\documentclass[a4paper,12pt]{article}
\usepackage{amsmath,amssymb}
\usepackage{graphicx,color}
\usepackage{soul}
\numberwithin{equation}{section}
\usepackage{cite}
\usepackage{bm}
\usepackage{dcolumn}
\newcommand{\be}{\begin{equation}}
\newcommand{\bea}{\begin{eqnarray}}
\newcommand{\eea}{\end{eqnarray}}
\newcommand{\ba}{\begin{array}}
\newcommand{\ea}{\end{array}}
\newcommand{\ee}{\end{equation}}

\expandafter\ifx\csname mathbbm\endcsname\relax

\else

\fi \textheight 22cm \textwidth 15cm \topmargin 1mm
\usepackage[T1]{fontenc} 
\usepackage{tikz}
\usetikzlibrary{arrows,snakes,backgrounds}
\usetikzlibrary{decorations.pathmorphing,backgrounds,shapes,arrows,shadows}
\tikzset{zigzag/.style={decorate,decoration=zigzag}}
\tikzset{snake it/.style={decorate, decoration=snake}}
\makeatletter
\def\@hex@@Hex#1%
 {\if a#1A\else \if b#1B\else \if c#1C\else \if d#1D\else
  \if e#1E\else \if f#1F\else #1\fi\fi\fi\fi\fi\fi \@hex@Hex}
\makeatother
\usepackage[all]{xy}
\usepackage[percent]{overpic}
\usepackage{slashed}
\usepackage{wrapfig}
\usepackage{tabu}
\usepackage{diagbox}
\usepackage{mathrsfs,amsmath,amssymb,amsthm,amsfonts,tikz,graphicx,accents,hyperref, color}
\usepackage{dsfont,epiolmec, latexsym, stmaryrd, comment}
\usepackage{slashed,ccaption}
\usepackage{mathrsfs, calligra}
\usepackage{nicefrac} 
\usepackage{xfrac} 
\usepackage{leftidx}
\usepackage{import}
\usepackage{multirow}
\usepackage{amsfonts}
\usepackage{pifont}
\usepackage{tabularx}
\usepackage{cancel}
\usepackage[utf8]{inputenc}
\usetikzlibrary{intersections,calc}
\usepackage{ifthen}
\usepackage{amsmath}
\usepackage{cancel}
\usepackage{caption}

\usepackage{array}
\hypersetup{ linktoc=all,
    colorlinks, linkcolor={blue},
    citecolor={red}, urlcolor={violet}
}

\graphicspath{{Images/}}

\def\sideremark#1{\ifvmode\leavevmode\fi\vadjust{\vbox to0pt{\vss
 \hbox to 0pt{\hskip\hsize\hskip1em
 \vbox{\hsize2cm\tiny\raggedright\pretolerance10000
 \noindent #1\hfill}\hss}\vbox to8pt{\vfil}\vss}}}%

\usepackage[most]{tcolorbox}

\tcbset{highlight math style={left=02mm,right=02mm,top=-1mm,bottom=-1mm}} 
\usepackage{empheq}

\DeclareSymbolFont{extraup}{U}{zavm}{m}{n}
\DeclareMathSymbol{\varheart}{\mathalpha}{extraup}{86}
\DeclareMathSymbol{\vardiamond}{\mathalpha}{extraup}{87}
\makeatletter
\renewcommand*{\@fnsymbol}[1]{\ensuremath{\ifcase#1\or \clubsuit \or \vardiamond \or \varheart\or
    \spadesuit\or \mathparagraph\or \|\or **\or \dagger\dagger
    \or \ddagger\ddagger \else\@ctrerr\fi}}
\makeatother
\definecolor{rosy}{RGB}{230,235,252}
\definecolor{myframetitle}{RGB}{90,89,170}
\definecolor{myblocktitle}{RGB}{140,185,249}
\definecolor{mytitle}{RGB}{10,80,26}

\definecolor{darkgreen}{RGB}{27,130,45}
\definecolor{darkblue}{rgb}{0,0,0.3}
\definecolor{darkred}{rgb}{0.7,0,0}

\definecolor{light gray}{RGB}{220,220,220}
\definecolor{dark purple}{RGB}{108,0,217}
\definecolor{pink}{RGB}{190,20,100}
\definecolor{orang}{RGB}{193,63,0}
\definecolor{green}{RGB}{11,98,17}
\definecolor{darkpink}{RGB}{153,0,76}
\definecolor{bluegreen}{RGB}{0,102,102}
\definecolor{greenlagan}{RGB}{0,102,0}
\definecolor{redgreen}{RGB}{102,102,0}
\definecolor{Redgreen}{RGB}{153,76,0}
\definecolor{vividviolet}{rgb}{0.62, 0.0, 1.0}
\definecolor{amaranth}{rgb}{0.9, 0.17, 0.31}
\definecolor{palatinateblue}{rgb}{0.15, 0.23, 0.89}
\definecolor{brightpink}{rgb}{1.0, 0.0, 0.5}
\definecolor{cornflowerblue}{rgb}{0.39, 0.58, 0.93}
\definecolor{deepcarminepink}{rgb}{0.94, 0.19, 0.22}
\definecolor{radicalred}{rgb}{1.0, 0.21, 0.37}

\newcommand\bc[1]{\boldsymbol{\mathcal{#1}}}

\newcommand{\bC}{{\bf C}}

\newcommand{\bapp}{\boldsymbol{\approx}}

\DeclareFontFamily{OT1}{rsfs}{}
\DeclareFontShape{OT1}{rsfs}{m}{n}{ <-7> rsfs5 <7-10> rsfs7 <10->rsfs10}{} 
\DeclareMathAlphabet{\mycal}{OT1}{rsfs}{m}{n}

\newcommand{\bcN}{\boldsymbol{\mathcal{N}}}
\newcommand{\bcO}{\boldsymbol{\mathcal{O}}}

\newcommand{\bL}{\mathbf{L}}

\newcommand{\bJ}{\mathbf{J}}
\newcommand{\bM}{\mathbf{M}}
\newcommand{\bK}{\mathbf{K}}

\makeatletter \@addtoreset{equation}{section}

\begin{document}
\begin{titlepage}
\begin{center}
 {\textbf{\LARGE{Towards Quantizing Null $p$-branes:}}}{\vskip 1mm {\Large{Light-Cone Gauge Analysis and Physical Hilbert Space}} \\ }
\vspace*{20mm}

{\large{S. Dutta$^a$, I. M. Rasulian$^b$, M.M. Sheikh-Jabbari$^b$, H. Yavartanoo$^c$ } }
\vspace*{1cm}

{\it ${}^a$ Physique th\'eorique et math\'ematique, Universit\'e Libre de Bruxelles, Campus Plaine - CP 231, 1050 Bruxelles, Belgium \\ }
  {\it ${}^b$ School of Physics, Institute for Research in Fundamental
Sciences (IPM), \\ P.O.Box 19395-5531, Tehran, Iran\\ }
{\it ${}^c$ Beijing Institute of Mathematical Sciences and Applications (BIMSA), Huairou District, \\ Beijing 101408, P. R. China \\ }
\vspace*{1cm}
\end{center}
\begin{abstract}
We study null $p$-branes, $p$-branes with a Carrollian $p+1$-dimensional worldvolume embedded in a generic $D$-dimensional flat Minkowski target space. This theory has a generalised BMS$_{p+1}$ gauge symmetry.  By fixing the light-cone gauge,  the BMS symmetry is partly fixed, leaving $p$ ``momentum constraints'' alongside $p$-dimensional area-preserving diffeomorphisms. We quantise the theory in the light-cone gauge via canonical quantisation and construct the physical Hilbert space by imposing the remaining constraints using sandwich conditions: the constraints must vanish when sandwiched between any two physical states.  We show that  solutions to the sandwich conditions are classified into $p+1$ distinct classes, which we completely specify. In addition, we  discuss special and interesting case of membranes in four dimensions and examine the physical implications of the quantised null $p$-brane and its associated  physical Hilbert space. 
\end{abstract}
\end{titlepage}

\tableofcontents 
\section{Introduction}

One of the early discoveries in string theory/supergravity was that the theory besides 1 dimensional strings or massless degrees of freedom associated with supergravity fields, also incorporates a variety of extended (solitonic) objects \cite{Duff:1994an} with different dimensions, known as $p$-branes \cite{Duff:1987cs}. In string theory $p$-branes dominate the dynamics when the theory is strongly coupled. String dualities  yielded to the notion of ``$p$-brane democracy'' \cite{Townsend:1995gp},  that there are regimes in the moduli space where  $p$-branes may be treated as ``fundamental objects'' and strings are non-perturbative degrees of freedom. So, it is equally important to study dynamics of $p$-branes as fundamental objects \cite{Polchinski:1998rq}. 

In its simplest form, a $p$-brane action is a $p+1$ dimensional extension of  strings worldsheet theory, which is essentially accounting for the $p+1$ dimensional area the brane probes as it moves in a given $D$ dimensional target space \cite{Duff:1987cs}. Thus, at classical level, {the} $p$-brane action is invariant under $p+1$ dimensional diffeomorphisms. For the $p=1$ case   (string theory)  this symmetry is further enhanced by the Weyl symmetry, {facilitating the} quantisation of the theory \cite{Green:1987sp, Polchinski:1998rq}. {However} for $p\geq 2$, in the absence of the Weyl symmetry, quantisation of $p$-branes faces {significant challenges}, most notably having  a continuous spectrum and the $p$-brane theory, even on flat target space, is non-renormalizable  \cite{deWit:1988xki}. Despite numerous efforts see e.g. \cite{Duff:1987cs, deWit:1989vb,Bytsenko:1992pw, Bergshoeff:1995hm, Russo:1996if, Russo:1996rw,Pavsic:1997eu, Yu:2011zza,Pavsic:2016icf},  quantisation of $p$-branes has remained an open problem.\footnote{For the case of membranes ($ p = 2 $), the problem in the light-cone gauge can be addressed via "matrix regularization," where the $ D-2 $ embedding coordinates of the membrane are promoted to $ N \times N $ matrices with $ N \gg 1 $ \cite{deWit:1988wri, Hoppe:1988gk, Taylor:2001vb}.  However, this approach encounters difficulties for higher-dimensional branes ($ p \geq 3 $) due to the challenge of "quantizing" Nambu $ p $-brackets \cite{Nambu:1973qe}. (Nambu 2-bracket corresponds to the standard Poisson bracket, and its quantisation is achieved by replacing the Poisson bracket with commutators.)  For attempts to quantise Nambu $ p $-brackets, see, e.g., \cite{Awata:1999dz, Hoppe:1996xp, Curtright:2002fd, Sheikh-Jabbari:2004fiz}.} In this work, we revisit and tackle this long-standing problem for {the} special case of null $p$-branes in a flat spacetime background,  providing a framework {that may} be extended to generic $p$-branes. 

Worldvolume of a null $p$-brane is by definition a $p+1$ dimensional null surface and as such null branes may be obtained as tensionless limit of usual $p$-branes \cite{Isberg:1993av, Hassani:1994rf}, as happens in a similar way for null strings \cite{Bagchi:2015nca,Bagchi:2016yyf, Bagchi:2020fpr, Bagchi:2021rfw, Bagchi:2019cay}. In the tensionless limit, the issues of a continuous spectrum and non-renormalisability that afflict conventional $ p $-branes are absent, making the analysis and quantisation of null branes more tractable. In the weakly coupled regime, where brane-brane interactions are perturbative, main challenge is consistent gauge-fixing and constructing the physical Hilbert space.  In other words, the equations of motion for null branes reduce to the null geodesic equation for the background spacetime, with a classical null brane configuration corresponding to a congruence of null geodesics.  The $p$-brane structure is {then} enforced through imposition of the constraints associated with $p+1$ dimensional diffeomorphisms on the worldvolume. The main question here is then how to consistently impose these constraints for a null brane theory at classical and quantum level. 

It has been argued that null strings/branes are the natural objects when the string/brane is probing horizon of a black hole \cite{Bagchi:2019cay, Bagchi:2020ats, Bagchi:2022iqb, Bagchi:2023cfp, Bagchi:2024rje, Bagchi:2024qsb}. Moreover, null strings/branes may be viewed as tensionless limit, $l_p\to\infty$ limit of usual tensile strings/branes where  $l_p$ is the fundamental length of the string/brane. Although this tensionless limit is diametrically opposite to the usual supergravity limit, it has also been a source of great intrigue for various considerations. In particular, because of the emergent simplicity of the string scattering amplitudes in this limit, initially observed by Gross and Mende \cite{Gross:1987kza,Gross:1987ar}. Tensionless limit is also relevant when we have a thermal gas of strings/branes at temperatures $T$, when $l_p T\gtrsim 1$ \cite{Pisarski:1982cn, Olesen:1985ej,Bagchi:2021ban, Bagchi:2020ats, Bagchi:2024tyq}. Worldvolume of null $p$-branes is a $p+1$ dimensional Carrollian space \cite{Henneaux:1979vn, Henneaux:2021yzg, Freidel:2022bai, Adami:2023wbe}. Null branes can be residing in a generic $D$ dimensional target space, which in this work and for simplicity, is taken  to be flat $D$ dimensional Minkowski space. Thus, geometrically, null brane worldvolume is embedding of a $p+1$ dimensional Carrollian space into a $D\geq p+2$ dimensional flat space. Null brane worldvolume theory in the temporal gauge enjoy  BMS$_{p+2}$ group extended by null Weyl scaling \cite{Grumiller:2019fmp, Adami:2020amw, Adami:2021nnf, Freidel:2021fxf} as the gauge symmetry.   

In this paper, we tackle the problem of quantising null $p$-branes in flat space in the light-cone gauge, where local symmetries of the brane theory {are} fixed up to $p$-dimensional translations plus $p$-dimensional area-preserving diffeomorphisms. {The} remaining/residual symmetries should be viewed as constraints. Our main point and observation is that consistency of the quantisation  only requires the constraints to vanish when sandwiched between any two physical states, analogous to recent approaches in null string theory \cite{Bagchi:2020fpr, Bagchi:2021rfw} and generic tensile strings \cite{Bagchi:2024tyq}.  We are able to solve these constraints and find the physical Hilbert space of the null $p$-brane theory in the light-cone gauge. We show there are $p+1$ classes of states in the Hilbert space, which we fully specify. 

This paper is organised as follows.  In section \ref{sec:null-brane-action}, we briefly review the classical action for a null $p$-brane, which may be obtained through a careful tensionless limit of the Polyakov-like action of a tensile $p$-brane. We discuss fixing the temporal gauge. In section \ref{sec:toridal-p-brane-LCG}, we focus on a toroidal null $p$-brane on a $D$ dimensional (flat) toroidal target space. We fix the light-cone gauge and perform canonical quantisation for the remaining/residual constraints and their algebra. In section \ref{sec:Hilbert-space}, we use ``sandwich quantisation'' proposal (outlined in appendix \ref{appen:Sandwich-quantisation}) to impose constraints,  classify all solutions to the sandwich constraints. We show solutions fall into $p+1$ classes, \textbf{Class} $\bcN$ $\bcN=0,\cdots, p$, and construct states in each class and hence the physical Hilbert space of the null $p$-brane theory. 
In section \ref{sec:p2D4} and to clarify more the general analyses and results of Section \ref{sec:Hilbert-space}, we focus on the specific example of a null membrane in four-dimensional spacetime ($p=2, D=4$ example). Section \ref{sec:discussion} is devoted to summary and discussions. In  appendix \ref{appen:Sandwich-quantisation} we review and briefly develop the general framework of  ``sandwich quantisation'', i.e. quantisation of constrained systems where the constraints are imposed through the requirement that a constraint should vanish when sandwiched between any two physical states. In  appendix \ref{sec:Categories-Class-II} we discuss completeness of $p+1$ classes of solutions for the null $p$-brane. 

\section{Null Brane Action}\label{sec:null-brane-action}

The action for a null $p$-brane in flat spacetime can be obtained as a certain tensionless limit of the standard  tensile  $p$-brane action on flat space, given by
\be\label{Brane-action-Polyakov}
S=-\frac{\tilde{T}}{2}\int d^{p+1}\sigma\ \sqrt{-h}\left(h^{ab}\partial_a X^\mu \partial_b X^\nu \eta_{\mu\nu}-\Lambda\right)
\ee
where $a, b=0,1,\cdots, p$ {and} $\sigma^a$ are the coordinates on the $p+1$ dimensional worldvolume. The fields $X^\mu$, $\mu=0,1,\cdots, D-1$ are embedding coordinates, $h_{ab}$ is the worldvolume metric with $h^{ab}$ as  its inverse, {and} $\eta_{\mu\nu}$  is the $ D $-dimensional Minkowski metric. The constant $ \Lambda $ appears as a Lagrange multiplier.  In the above action dynamical fields are $h_{ab}$ and $X^\mu=X^\mu(\sigma)$.  Our analysis  keeps  $p, D$ general. In {the} conventions {used here} (in accord with  standard string theory conventions \cite{Green:1987sp, Polchinski:1998rq}),  {the coordinates} $\sigma^a$ {are}  dimensionless, $X^\mu$ have  dimension of length, $\Lambda$ has dimension of length$^2$, $h_{ab}$ {is} dimensionless, and hence $\tilde{T}$ has dimension of length$^{-2}$.

Action \eqref{Brane-action-Polyakov} is invariant under $p+1$ dimensional {worldvolume diffeomorphisms} and under  scaling transformations of the metric, $h_{ab}\to W h_{ab}$, 
$$
\sqrt{-h}\left(h^{ab}\partial_a X^\mu \partial_b X^\nu \eta_{\mu\nu}-\Lambda\right) \to W^{\frac{p-1}{2}} \sqrt{-h}\left(h^{ab}\partial_a X^\mu \partial_b X^\nu g_{\mu\nu}-W\Lambda\right).
$$
{Thus, Weyl invariance is retained only for $ p = 1 $ and $ \Lambda = 0 $.}  For convenience we denote the induced metric on the brane by $G_{ab}$:
$$
G_{ab}:= \partial_a X^\mu \partial_b X^\nu \eta_{\mu\nu}.
$$
{The} equations of motion for $h_{ab}$ {derived from \eqref{Brane-action-Polyakov} } are
\be\label{hab-EoM}
h_{ab}=\frac{p-1}{\Lambda} G_{ab} \quad \text{for}\quad p\neq 1, \quad \Lambda=0 \quad \text{for}\quad p=1.
\ee
{Substituting} $h_{ab}$ into the action \eqref{Brane-action-Polyakov} for $p\neq 1$ we obtain the Nambu-Goto type $p$-brane action
\be\label{Brane-action-NG-type}
S=-T \int d^{p+1}\sigma\ \sqrt{-\deg{G}}, \qquad T=(p+1)\left(\frac{1-p}{\Lambda}\right)^{\frac{p-1}{2}}\tilde{T}.
\ee

\paragraph{Null brane limit.} Let us denote $\sigma^a=(\tau,\sigma^i), i=1,2,\cdots,p$ and parameterise the worldvolume metric $h_{ab}$ as
\be\label{null-h}
h_{\tau\tau}=-\epsilon^2 (\frac1{{\lambda}^2}-U^2),\qquad h_{\tau i}=h_{i \tau}=\epsilon h_{ij}U^j, \qquad U^2:=h_{ij}U^iU^j
\ee
and $h_{ij}\sim {\cal O}(1)$ while $\epsilon\to 0$. For this case, 
\be
\begin{split}
\sqrt{-h}:= &\sqrt{-\det{h_{ab}}}=  \frac{\epsilon}{\lambda} \sqrt{\det{h_{ij}}} ,  \\
h^{\tau\tau}=-\frac{{\lambda}^2}{\epsilon^2},\qquad h^{\tau i}&=\frac{{\lambda}^2}{\epsilon} {U^i},\qquad h^{ij}=(h^{-1})^{ij}-{{\lambda}^2}{U^i U^j}
\end{split}
\ee
where $(h^{-1})^{ij} h_{jk}=\delta^i{}_k$. In this limit worldvolume geometry of the $p$-brane reduces to a Carrollian geometry (see e.g. \cite{Freidel:2022bai, Adami:2023wbe}) with a kernel vector along  $\partial_\tau$ and  a $p$-dimensional metric $h_{ij}$. The next step  is to explore the equations of motion and constraints in the above limit. However,  before doing so, we note that the null brane limit can be defined more generally: instead of requiring the norm of $\partial_\tau$ to vanish as $\epsilon^2$, we could take the limit such that the norm of an arbitrary  vector $V^a$ approaches  to zero as $\epsilon^2$. Explicitly, instead of \eqref{null-h} we can define the limit via
\begin{equation}\label{V-kernel-limit}
h_{ab}V^aV^b\sim\epsilon^2\to 0,\quad h_{ab}V^a X^b\sim \epsilon\to 0,\quad h_{ab}X^a X^b\sim {\cal O}(1), 
\end{equation}
where $X^a$ is a generic vector not aligned with $V^a$. In this way, we obtain a Carrollian worldvolume with a kernel along $V^a$ and a $p$-dimensional metric $h_{ij}$. Note that \eqref{V-kernel-limit} remains invariant under rescaling $V^a$ by an arbitary function. We will discuss this emergent symmetry further later  in  this section. To simplify the analysis in the following discussion, we take $V^a\partial_a \propto \partial_\tau$.

We begin by exploring the  equations of motion  in the Carrollian limit described above.  For tensile $p$-branes, the equations of motion are given by
\begin{align} \label{tensileeom}
	\partial_a (\sqrt{-h}h^{ab}\partial_b X^{\mu})=0.
\end{align}
In the null limit and at the leading order, this reduces to
\begin{align}\label{EoM-limit}
	\partial_{\tau}(\lambda \sqrt{\det h_{ij}}\partial_{\tau} X^\mu)=0\,.
\end{align}
The gauge freedom (rescaling of the kernel vector) allows us to set $\lambda\sqrt{\det{h_{ij}}}=1$. We will revisit  this equation in the next section.

Next, we examine the constraints in the null brane limit. The constraints, equations of motion for $h_{ab}$ \eqref{hab-EoM}, reduce in the limit to,
\begin{equation}\label{constraint-null}
    G_{\tau\tau}\sim \epsilon^2 \to 0, \qquad  G_{i\tau}\sim \epsilon \to 0, \qquad G_{ij}= {\cal X} h_{ij}
\end{equation}
where ${\cal X}$ is an arbitrary constant. In the final paragraphs of this section,  we will comment on the physical interpretation of this arbitrary constant. The constraints in the limit are hence $G_{\tau\tau}=0, G_{\tau i}=0$, that is, 
\begin{equation}\label{constraints-limit}
C_0:=\partial_{\tau} X^{\mu} \partial_{\tau} X_{\mu}=0,\qquad C_i:=\partial_{\tau} X^{\mu} \partial_{i} X_{\mu}=0.
\end{equation}

One may investigate whether \eqref{EoM-limit} and \eqref{constraints-limit} can be obtained from an action principal. To examine this possibility, we consider the null worldvolume limit of the brane action \eqref{Brane-action-Polyakov}. This yields: 
\be\label{brane-action-epsilon}
S=\frac{\kappa}{2}\int d\tau d^{p}\sigma\ {{\lambda}}\ \sqrt{\det{h_{ij}}}\left[(\partial_\tau X^\mu-\epsilon U^i\partial_i X^\mu)^2-\frac{\epsilon^2}{{\lambda}^2}\left( (h^{-1})^{ij}\partial_i X^\mu\partial_j X^\nu \eta_{\mu\nu}+\Lambda\right)\right]
\ee
where $\kappa=\tilde{T}/\epsilon$. Taking the $\epsilon \to 0$ limit (the null brane limit) while keeping $\kappa$ and other parameters in the action fixed, we obtain:
\be\label{Null-brane-action-tau-gauge}
S_{\text{N.B.}}=\frac{\kappa}{2}\int d\tau \int_{{\cal S}_p}d^{p}\sigma\ \sqrt{\det{{h}_{ij}}}\ \left[\lambda\ \partial_\tau X^\mu\partial_\tau X^\nu \eta_{\mu\nu}\right]
\ee
Here, the ``null $p$-brane tension'' $\kappa$, has  dimensions of length$^2$, the same as $\tilde{T}$. We note that in keeping the action finite, we need to send the brane tension $\tilde{T}\sim \epsilon\to 0$. That is, a null brane limit is a certain tensionless limit of the original tensile brane. Varying this action with respect to $X^\mu$ reproduces the equation of motion \eqref{EoM-limit}, while variation with respect to $\lambda$ yields the constraint $C_0=0$ from \eqref{constraints-limit}. To derive the remaining constraints $C_i=0$  from the action, we must generalize the kernel vector to $V^a$ as introduced in \eqref{V-kernel-limit} and the surrounding discussion. Specifically, replacing $\partial_\tau$ with  $V^a\partial_a$ leads to the null brane action

\be
S_{\text{N.B.}}=\frac{\kappa}{2}\int_{{\cal N}_{p+1}}d^{p+1}\sigma\ \lambda \sqrt{\det{h_{ij}}}\ V^a\partial_a X^\mu\ V^b\partial_b X^\nu \eta_{\mu\nu}
\ee
where ${\cal N}_{p+1}$ denotes the $p+1$ dimensional worldvolume Carrollian  geometry specified by $(V^a, h_{ij})$ (see e.g. \cite{Freidel:2022bai, Adami:2023wbe}). In taking the null brane limit we chose a specific time direction $\tau$ (which becomes null in the limit) and the vector $V^a$, besides reproducing the constraint equations \eqref{constraint-null}, retains {the} $p+1$ dimensional diffeomorphism invariance of the action. The  null brane action  may also be written in the standard ILST form  \cite{Isberg:1993av, Hassani:1994rf},
\be\label{null-brane-action-covariant}
\tcbset{fonttitle=\scriptsize}
            \tcboxmath[colback=white,colframe=gray]{
S_{\text{N.B.}}=\frac{\kappa}{2}\ \int_{{\cal N}_{p+1}}d^{p+1}\sigma\  {\cal V}^a\partial_a X^\mu\ {\cal V}^b\partial_b X^\nu \eta_{\mu\nu}}
\ee
where ${\cal V}^a$ is a vector density,
\be\label{cal-V-density}
{\cal V}^a :=  \lambda^{1/2}(\det{h_{ij}})^{1/4} V^a.
\ee

Variation of the null brane action yields
\bea \label{ILST}
\delta S_{\text{N.B.}}&=&{\kappa}  \int_{{\cal N}_{p+1}}  \left( {\cal V}^a\partial_a X^\mu\ \partial_b X^\nu \eta_{\mu\nu}\right) \delta {\cal V}^b - \partial_b ({\cal V}^a {\cal V}^b\partial_a X^\mu\  \eta_{\mu\nu})\delta X^\nu \nonumber \\ & + & {\kappa}\int_{\partial{\cal N}_{p+1}} n_a{\cal V}^a \ {\cal V}^b \partial_b X^\mu\  \eta_{\mu\nu} \delta X^\nu ,
\eea
where $n_a$ is {a vector field} transverse to the worldvolume ${\cal N}_{p+1}$. Equations of motion corresponding to $X^\mu$ and  $\mathcal{V}^a$ are respectively given by
\begin{subequations}
\begin{align}
\partial_a( & {\cal V}^a\  {\cal V}^b \partial_b X^\mu\  \eta_{\mu\nu}) =0 \label{EoM}\\
{\cal C}_a(\tau, \sigma^i)&:={\cal V}^a\partial_a X^\mu\ \partial_b X^\nu \eta_{\mu\nu} =0 \label{Constraints-Va},
\end{align}
\end{subequations}
\eqref{Constraints-Va}, which {are} a rewriting of \eqref{constraint-null},  are constraints {on} the field equations \eqref{EoM}. 

Vanishing of $\int_{\partial({\cal S}_p\times {\mathbb R})} n_a{\cal V}^a \ {\cal V}^b \partial_b X^\mu\  \eta_{\mu\nu} \delta X^\nu$  yields boundary conditions. We will explore these equations in the following sections. 


\subsection{Symmetries of the Null Brane Action}

The original action \eqref{Brane-action-Polyakov} is invariant under local $p+1$ dimensional worldvolume diffeomorphisms  along with global Poincar\'e symmetry of the target space.  This invariance can be verified by considering infinitesimal diffeomorphisms generated by  $\xi^a(\sigma)$, with the following transformations 
\be \begin{split}
\delta_\xi X^\mu = {\mathcal L}_\xi X^\mu = \xi\cdot \partial X^\mu &,\qquad  \delta_\xi \eta_{\mu\nu} = {\mathcal L}_\xi \eta_{\mu\nu} = \partial_\mu\xi_{\nu}+ \partial_\nu\xi_{\mu}\\
\delta_\xi h_{ij} = {\mathcal L}_\xi h_{ij} = \nabla_i\xi_j +\nabla_j \xi_i &, \qquad \delta_\xi \lambda =\partial_\tau\xi^\tau\lambda\\
\delta_\xi V^a={\mathcal L}_\xi V^a=
\xi\cdot \partial V^a-V^b \partial_b \xi^a  &,\qquad \delta_\xi {\cal V}^a={\mathcal L}_\xi {\cal V}^a=
\xi\cdot \partial {\cal V}^a-{\cal V}^b \partial_b \xi^a -\frac{1}{2}(\partial_b\xi^b) {\cal V}^a\ .
\end{split}\ee 
Under these transformations, the action remains invariant.

Crucially, the null brane action \eqref{null-brane-action-covariant} possesses an additional  local symmetry not shared by the tensile brane action. {This is an extension of the arbitrary constant ${\cal X}$ appeared in \eqref{constraint-null}.} {Specifically,} it is invariant under the null Weyl scaling Weyl$_{{\cal N}}$:
\be\label{null-Weyl}
\lambda\to\lambda,\qquad
V^a\to W^{-1/2} V^a, \qquad h_{ij}\to W^{2/p} h_{ij}, \qquad W=W(\tau, \sigma^i).
\ee
{Notably,} the vector density ${\cal V}^a$ {remains} invariant under the  Weyl$_{{\cal N}}$ scaling. Thus the full symmetry of null $p$-brane theory is ${\cal A}_{\text{N.B.}}:=$ Diff$(p+1)\inplus$ Weyl$_{{\cal N}}$, where  ${\cal A}_{\text{N.B.}}$ represents the  null boundary symmetry algebra,  discussed in \cite{Adami:2021nnf, Freidel:2021fxf}. In summary, the null brane theory \eqref{null-brane-action-covariant} {is characterised by} ${\cal A}_{\text{N.B.}}$ as its local (gauge) symmetry and it has the right symmetries to be {a strong} candidate for the theory describing boundary degrees of freedom residing at a null boundary. The null boundary algebra ${\cal A}_{\text{N.B.}}$ has a direct geometric interpretation as the symmetry group of the worldvolume Carrollian geometry.

\subsection{Temporal Gauge Fixing}\label{sec:p-brane-EoM-null-congruence}

As in the  case of tensile brane, studying  {the null brane theory requires} fixing the diffeomorphism invariance. The worldvolume vector ${\cal V}$ can be fixed in the ``temporal gauge'' leaving us with  foliation-preserving subset of the diffeomorphisms, explicitly,  
\be\label{temportal-gauge-fixing}
{\cal V}^a \partial_a =\partial_\tau
\ee
In this gauge the induced metric on the spatial slices becomes $h_{ij}=\partial_i X^\mu \partial_j X^\nu \eta_{\mu\nu}$,  and the action takes the form:  
\be
S=\frac{\kappa}{2}\int d\tau \int_{{\cal S}_p}d^{p}\sigma\ \partial_\tau X^\mu\ \partial_\tau X^\nu \eta_{\mu\nu}.
\ee
The equations of motion derived from this action are:
\be\label{NB-EoM}
\partial_\tau (\eta_{\mu\nu} \partial_\tau X^\nu)=0. 
\ee
The above equation describes affinely parametrized null geodesics over target space. {The solutions represent} a congruence of massless particles uniformly distributed over {the} $p$-dimensional spacelike surface ${\cal S}_p$. However, this congruence describes  a null brane once the congruence is subjected to {the following} $p+1$ constraints \eqref{Constraints-Va}
\be\begin{split} \label{temp eom}
\partial_\tau X^\mu \partial_a X^\nu \eta_{\mu\nu}=0. 
\end{split}
\ee
With this equation in place, it would be convenient to introduce the following constraints:
\be\label{hab-EoM-null}
\begin{split}
{\cal C}_0(\tau, \sigma^i)= \partial_\tau X^\mu \partial_\tau X^\nu \eta_{\mu\nu},\qquad {\cal C}_i(\tau, \sigma^i)=\partial_\tau X^\mu \partial_i X^\nu \eta_{\mu\nu} 
\end{split}
\ee

It is important to  note that the null brane equations of motion \eqref{NB-EoM} are linear in embedding coordinates $X^\mu$, which makes them straightforward to solve. This is to be compared with the tensile $p$-brane case where equations of motion involve higher order terms (typically terms with  $\sim X^{2p}$). Using equations of motion we obtain
\be
\partial_\tau {\mathcal  C}_0=0,\qquad  
\partial_\tau\left({\mathcal C}_i -\frac{1}{2}\;\tau  \partial_i {\mathcal C}_0\right)=0
\ee
Thus, ${\mathcal  C}_0$ is $\tau$-independent, and ${\cal C}_i={\cal C}_i^0+\frac12\tau \partial_i{\mathcal  C}_0$, where ${\cal C}_i^0$ is also $\tau$-independent. Therefore, the system is effectively subject to  two $\tau$-independent constraints: ${\cal C}_i^0=0$ and ${\cal C}_0=0$. This is in contrast to the tensile brane case, where the constraints are  generally $\tau$-dependent on-shell, requiring additional secondary constraints to ensure that the time derivatives of the constraints vanish. Consequently, the null brane theory has half as many constraints as the tensile case.

In the temporal gauge $\mathcal V=\partial_\tau$, the boundary term of the action is
\be 
\kappa \int d^p\sigma\  \partial_\tau X^\mu\   \delta X_\mu\bigg\rvert_{\tau=\tau_i}^{\tau=\tau_f}=0\quad 
\ee 
Thus, the boundary  for the null brane is a constant-time slice of  the worldvolume,  setting the initial conditions. This boundary term may be contrasted with the tensile brane case, where the boundary  involves time-like sectors and yields non-trivial boundary conditions.  For a compact, closed  ${{\cal S}_p}$ with ``periodic boundary conditions'', the above boundary term vanishes on-shell. In this work, we focus on closed $p$-branes with ${{\cal S}_p}$ being a $p$ dimensional torus $T^p$, satisfying such periodic boundary conditions.

\paragraph{Residual symmetries.} The gauge symmetries of the null brane theory,  Diff$(p+1)\inplus$ Weyl$_{{\cal N}}$, are generated by {arbitrary} functions of $\tau$ and $\sigma^i$. {Imposing}  the temporal gauge fixes $\tau$-dependence of the symmetry generators, leaving residual foliation-preserving (gauge) symmetries. With temporal gauge fixing, the residual  symmetry generators (diffeomorphisms) take the form:
\be\label{trans}
\zeta=\big(\sum_I \partial_i f^i(\sigma^j)\tau+h(\sigma^j)\big)\partial_\tau+\sum_I f^i(\sigma^j)\partial_i, 
\ee
where the null brane coordinates are $\tau$ and $\sigma^i$ (with $i$ from 1 to $p$). The residual symmetry generators are parametrised by functions $f^i(\sigma^j)$ and $h(\sigma^j)$. 
We  define the operators $L^{(i)}(f^i)$ (with no sum over $i$) and $M(h)$ as follows:
\be
\begin{split}
L^{(i)}(f^i)&=f^i \partial_i+(\partial_i f^i)\tau \partial_\tau,\\
M(h)&=h \partial_\tau,
\end{split}
\ee
again with no sum over $i$. The commutation relations (Lie brackets) between these operators are given by:
\be
\begin{split}
[L^{(i)}(f^i),L^{(j)}(g^j)]&=f^i L^{(j)}(\partial_i g^j)-g^j L^{(i)}(\partial_j f^i) \\
[L^{(i)}(f^i),M(h)]&=M(f^i \partial_i h-h \partial_i f^i),
\end{split}
\ee
The above is a BMS$_{p+2}$ algebra discussed in \cite{Grumiller:2019fmp}, it is an extension of  Compiglia-Laddha extended BMS$_4$ \cite{Campiglia:2015yka} to arbitrary dimensions. For $p=1$ this is BMS$_3$ symmetry of the null string \cite{Bagchi:2013bga}.

Besides the residual BMS$_{p+2}$ diffeomorphisms, the Weyl$_{{\cal N}}$ symmetry remains intact, as $\mathcal V^a$ is invariant under $W$ scaling. However, as we will be shown below, this Weyl symmetry in the null brane theory does not yield an independent symmetry generator/constraint on the system.

\subsection{Canonical Computation Relations and Algebra of Constraints} 

The conjugate momentum to $X^\mu$ in this gauge is given by
\be\label{can-momentum}
P_\mu= \kappa \eta_{\mu\nu} \partial_\tau X^\nu,\qquad P^\mu=\kappa \partial_\tau X^\mu
\ee
A generic null geodesic of the target space metric $\eta_{\mu\nu}$ can be denoted by $\gamma(x^\mu_0, p^\mu_0)$, where $x^\mu_0, p^\mu_0$ are initial position and momentum satisfying $p_0^\mu p_0^\nu \eta_{\mu\nu}=0$. This path is parametrised by $\tau$ and is given by $X^\mu_\gamma(\tau)$, such that
\be\label{x0p0}
X^\mu_\gamma(\tau=0)=x_0^\mu, \qquad P^\mu{}_{\gamma}(\tau=0)=p_0^\mu=\kappa \partial_\tau X^\mu_\gamma.
\ee
The most general solution to the equations of motion \eqref{NB-EoM} can then be expressed as,
\be\label{NB-Solution}
X^\mu(\tau; \sigma_i)= X^\mu_{\gamma(x_0^\mu(\sigma_i), p_0^\mu(\sigma_i))}(\tau),
\ee
where $x_0^\mu(\sigma_i)$ and $p_0^\mu(\sigma_i)$ are arbitrary functions over ${\cal S}_p$. We remark that, as \eqref{NB-Solution} indicates, solutions to null $p$-brane equations of motion are essentially a congruence of null geodesics labelled by $\gamma(x_0^\mu(\sigma_i), p_0^\mu(\sigma_i))$. 

The congruence of null geodesics behaves as a null $p$-brane upon imposing ${\cal C}^a=0$ constraints \eqref{hab-EoM-null}. The null $p$-brane is particularly interesting because, unlike the tensile $p$-brane case, its equations of motion can be explicitly solved, leaving the imposition of constraints-especially at the quantum level-as the primary challenge. This is the main focus of our work.  

\paragraph{Algebra of constraints.} Recalling the canonical Poisson brackets, 
\be\label{canonical-Poisson-brackets}
\{ X^\mu(\tau, \sigma^i), P_\nu(\tau, \tilde{\sigma}^i)\}= \delta^\mu_\nu\ \delta^p(\sigma-\tilde{\sigma})
\ee
One can compute the algebra of constraints ${\cal C}_0(\tau,\sigma^i)$ and ${\cal C}_i(\tau,\sigma^i)$ \eqref{hab-EoM-null}:
\be\label{algebra-constraints}
\begin{split}
    \{ {\cal C}_0 (\tau, \sigma^i), {\cal C}_0(\tau, \tilde{\sigma}^i)\}&=0\\
    \{ {\cal C}_0 (\tau, \sigma^i), {\cal C}_i(\tau, \tilde{\sigma}^i)\}&= \left[ {\cal C}_0 (\tau, \tilde{\sigma}^i)\frac{\partial}{\partial\sigma^i}-{\cal C}_0 (\tau, \sigma^i)\frac{\partial}{\partial\tilde{\sigma}^i}\right] \delta^p(\sigma-\tilde{\sigma})\\
    \{ {\cal C}_i (\tau, \sigma^i), {\cal C}_j(\tau, \tilde{\sigma}^i)\}&= \left[{\cal C}_i (\tau, \tilde\sigma^i)\frac{\partial}{\partial\sigma^j} -  {\cal C}_j(\tau, {\sigma}^i)\frac{\partial}{\partial\tilde{\sigma}^i}\right]\delta^p(\sigma-\tilde{\sigma})
\end{split}
\ee
Here ${\cal C}_0$ is generator of supertranslation part of BMS$_{p+2}$  and ${\cal C}_i$ generates diffeomorphisms on ${\cal S}_p$. The null $p$-brane Lagrangian is just ${\cal C}_0$. The null brane configurations should fall into the ${\cal C}_0=0={\cal C}_i$ representations (coadjoint orbits) of this algebra. Consider ${\cal W}(\tau,\sigma^i):= X^\mu(\tau,\sigma)P_\mu(\tau,\sigma)$. It follows that $\partial_\tau {\cal W}(\tau,\sigma^i)= {\cal C}_0+ X^\mu(\tau,\sigma)\partial_\tau P_\mu(\tau,\sigma)$ which vanishes on-shell when the constraint ${\cal C}_0=0$ is imposed. The Poisson brackets involving \({\cal W}\) are: 
\be\label{W-algebra-constraints}
\begin{split}
     \{ {\cal W} (\tau, \sigma^i), {\cal W}(\tau, \tilde{\sigma}^i)\}&=0\\
    \{ {\cal W} (\tau, \sigma^i), {\cal C}_0(\tau, \tilde{\sigma}^i)\}&= 2{\cal C}_0(\tau, {\sigma}^i)  \delta^p(\sigma-\tilde{\sigma})\\
    \{ {\cal W} (\tau, \sigma^i), {\cal C}_i(\tau, \tilde{\sigma}^i)\}&=  
    - {\cal W} (\tau, \tilde\sigma^i) \frac{\partial}{\partial\sigma^i}
    \delta^p(\sigma-\tilde{\sigma})
    \end{split}
\ee

The algebra of constraints of the null brane, the algebra generated by ${\cal C}_a$ and ${\cal W}$, is equivalent to the local gauge symmetry of the null brane theory:   Diff$(p+1)\inplus$ Weyl$_{{\cal N}}$. As any gauge theory, fixing the gauge symmetry is consistent only when the corresponding constraints are imposed. Finally, the null brane action can be written in terms of explicitly  Weyl$_{{\cal N}}$-invariant quantities using ${\cal V}^a$ \eqref{cal-V-density}. This confirms that the Weyl symmetry generator does not introduce any additional independent constraint, and ${\cal W}(\tau,\sigma^i)$ remains constant on-shell.

\section{Toroidal Null \texorpdfstring{$p$-brane}{p-brane} in Light-Cone Gauge}\label{sec:toridal-p-brane-LCG}
To proceed, we supplement the temporal gauge condition \eqref{temportal-gauge-fixing} with solving a part of the constraints \eqref{hab-EoM-null}. In this section, we demonstrate  how this can be achieved in the  light-cone gauge. 

We consider a toroidal $p$-brane in a $D$ dimensional target spacetime, i.e. $\mu=0,1,\cdots, D-1$ and $i=1,2,\cdots, p$. The solutions to  the equations of motion  can be expanded  as follows:
\be \label{Xexpansion}
X^\mu=x_0^\mu + A_i ^\mu \sigma^i +  \frac{1}{\kappa} B^\mu_0 \tau +\sum_{\vec{n}\neq \vec{0}} \left(\frac{1}{|\vec{n}|}\ A^\mu_{\vec{n}}\ e^{-i\vec{n}.\vec{\sigma}}+ \frac{1}{\kappa}B^\mu_{\vec{n}} \tau\ e^{-i\vec{n}.\vec{\sigma}}\right), \qquad \mu\in\{+, -, I\},
\ee
where $I=1,2,\cdots D-2$, $\vec{n}$ is a $p$-vector with integer-valued components $n_i$ and $|\vec{n}|=\sqrt{\sum_{i=1}^p n_i^2}$.  The reality condition for $X^\mu$ necessitates: 
\begin{equation}
(A^\mu_{\vec{n}})^*=A^\mu_{-\vec{n}}, (B^\mu_{\vec{n}})^*=B^\mu_{-\vec{n}}.     
\end{equation}

The constraints in the temporal gauge {can} be expressed through the Fourier modes of $L^i, M$ {as} defined in the previous section, i.e.  
\be 
L^i_{\vec{n}}:=\int d^p \sigma\ ({\cal C}_i+i n_i \tau {\cal C}_0)e^{i\vec{n}.\vec{\sigma}}= \int d^p \sigma \left(\partial_\tau X^\mu  \partial_i X_\mu + \frac{1}{2} i n_i \tau \partial_\tau X^\mu  \partial_\tau X_\mu\right) e^{i\vec{n}.\vec{\sigma}},
\ee
and 
\be 
M_{\vec{n}}:=\int d^p \sigma\ {\cal C}_0\ e^{i\vec{n}.\vec{\sigma}} =\int d^p \sigma \left( \frac{1}{2}  \partial_\tau X^\mu  \partial_\tau X_\mu\right) e^{i\vec{n}.\vec{\sigma}}. 
\ee
Substituting the expansion \eqref{Xexpansion}, these become:  
\be
M_{\vec{n}}= \frac{(2\pi)^p}{2} \sum_{\vec{k}} B_{\mu\vec{k}} B^{\mu}_{\vec{n}-\vec{k}},\qquad L^i_{\vec{n}}=\frac{(2\pi)^p}{2} \big( 2B_{\vec{n}\mu} A_i^\mu-i\sum_{\vec{k}\neq0} \frac{k_i}{|\vec{k}|} A_{\vec{k}\mu} B_{\vec{n}-\vec{k}}^\mu\big),
\ee 
{where} $M_{\vec{n}}$ {and} $L^i_{\vec{n}}$ are $\tau$-independent. 
The constraints reduce to the following relations for the mode expansion coefficients:
\begin{subequations}
\begin{align}
B_0^\mu B_{0\mu}+ \sum_{\vec{k}\neq 0}B_{\mu \vec{k}}B^\mu_{-\vec{k}}&=0, \label{momentum-mass-1}\\
2B_0^\mu B_{\vec{n}\mu}+ \sum_{\vec{k} \neq \{0,\vec{n}\} }B_{\mu \vec{k}}B^\mu_{\vec{n}-\vec{k}}&=0, \qquad \vec{n}\neq 0 \label{B-n-1}\\
B^\mu_{0} A_{\mu i} -i\sum_{\vec{k}\neq 0} \frac{ k_i}{|\vec{k}|}\ B^\mu_{-\vec{k}} A^\mu_{\vec{k}} &=0, \label{level-matching-1}\\
\frac{n_i}{|\vec{n}|} B^\mu_0 A^\mu_{\vec{n}}+i B^\mu_{\vec{n}} A_{\mu i} +\sum_{\vec{k}\neq \{0,\vec{n}\}} \frac{k_i}{|\vec{k}|}\ B^\mu_{\vec{n}-\vec{k}} A^\mu_{\vec{k}} &=0,\qquad \vec{n}\neq 0 \label{A-n-1}
\end{align}\end{subequations}
where \eqref{level-matching-1} and \eqref{A-n-1} should be satisfied for all $i=1,\cdots,p$.

\subsection{Light-cone Gauge Fixing}\label{sec:LCG-fixing} 

Solutions to the equations of motion in the temporal gauge are specified through $2(p+1)$ functions of $\sigma^i$, whose Fourier modes are denoted as  $A^\mu_{\vec{n}}, B^\mu_{\vec{n}}$.  These {functions} are subject to constraints ${\cal C}_a$, which  give rise to $p+1$ constraints $M$ and  $L^i$, as functions of $\sigma^i$. We can explicitly solve some  of these constraints by fixing the light-cone gauge. We start with  adopting the target space coordinates system $X^+, X^-, X^I$.  In the light-cone gauge we set $A^+_{\vec{n}}=B^+_{\vec{n}}=0$, i.e.  
\be 
X^+=x_0^+ + {p^+} \tau.
\ee
For simplicity we  assume the torus to be an orthogonal torus with radii $R_I$, then the winding and momentum modes are given by 
\be\label{winding-momentum}
{B_0^+={p^+}},  \qquad {B_{\vec{n}}^+=0} , \qquad  B^I_0= \frac{m_I}{R_I}, \qquad A^I_j= w^I_j R_I,\qquad  A^\pm_i=0, \qquad w^I_j, m_I, \in \mathbb{Z}. 
\ee

The constraints then determine $X^-(\tau, \sigma^i)$. Specifically,  \eqref{momentum-mass-1} and \eqref{level-matching-1} yield the following:
\begin{subequations}
\begin{align}
M^2 &:=2 p^+ B^-_{0}=\sum_I\frac{m^2_I}{R_I^2}+ \frac12\sum_{\vec{k}\neq 0}\ B^{{I}}_{\vec{k}}B^{{I}}_{-\vec{k}}, \label{momentum-mass}\\
L_i &:= m_I w^I_i-i\sum_{\vec{k}\neq 0}\ \frac{k_i}{|\vec{k}|} A^{{I}}_{\vec{k}} B^{{I}}_{-\vec{k}}=0,\qquad \text{no sum over} \ i. \label{level-matching}
\end{align}\end{subequations}
Eqs.~\eqref{B-n-1} and \eqref{A-n-1} {further determine} $B^-_{\vec{n}}$ and $A^-_{\vec{n}}$ in terms of transverse modes:
\begin{subequations}
\begin{align}
p^+ B^-_{\vec{n}}&= \frac{m_I}{R_I} B^I_{\vec{n}}+{\frac1{2}}\sum_{\vec{k}\neq 0,\vec{n}} \ B^{{I}}_{\vec{k}} B^{{I}}_{\vec{n}-\vec{k}} \label{B-n}\\ 
\frac{n_i}{|\vec{n}|}  p^+ A^-_{\vec{n}} &=i w^I_i R_I B^I_{\vec{n}}+ \frac{n_i}{|\vec{n}|}\sum_I \frac{m_I}{R_I} {A}^{{I}}_{\vec{n}} + \sum_{\vec{k}\neq 0,\vec{n}} \frac{k_i}{|\vec{k}|}\ A^{{I}}_{\vec{k}} B^{{I}}_{\vec{n}-\vec{k}},\qquad \text{no sum over} \ i. \label{A-n}
\end{align}\end{subequations}
where $i=1,\cdots, p$. 

Eq.~\eqref{momentum-mass} specifies the light-cone Hamiltonian $B^-$, associated with configurations that may be interpreted as a light-cone mass \footnote{This should not be confused with the physical mass of the membrane in the target space. It's important to note that the world-volume of the membrane is a null surface, meaning that, roughly speaking, it is an object moving with the speed of light, and it is considered a massless object from the target space point of view. Also, there are more subtleties in defining the mass in the quantised theory. See section \ref{sec:mass} for more discussions.} and \eqref{level-matching} is counterpart of level matching for the null brane case, for a $p$-brane we have $p$ level matching conditions. 
Eq.~\eqref{B-n} specifies $B^-_{\vec{n}}$, while \eqref{A-n} consist of $p$ equations for a single unknown $A^-_{\vec{n}}$ and requires further analysis. Multiplying both sides  of \eqref{A-n} by $\frac{n_i}{|\vec{n}|}$ and summing over $i$ we get,
\begin{equation}\label{A-n-parallel}
     p^+ A^-_{\vec{n}} =i \sum_{j,I} \frac{n_j}{|\vec{n}|}w^I_j R_I B^I_{\vec{n}}+\sum_J \frac{m_J}{R_J} {A}^{{J}}_{\vec{n}} +  \sum_{\vec{k}\neq 0,\vec{n}} \frac{\vec{n}\cdot\vec{k}}{|\vec{n}||\vec{k}|}\ A^{{J}}_{\vec{k}} B^{{J}}_{\vec{n}-\vec{k}},
\end{equation}
which resembles the equation for $B^-_{\vec{n}}$ in \eqref{B-n}. 

Next we  project both sides of \eqref{A-n} onto directions transverse to $\vec{n}$. To this end, we multiply both sides by the projector $(T^r_{\vec{n}})^{i}$, {which}  projects out the part of $X_i$ parallel to $\vec{n}$. So in general it has $r=1,2,\cdots, p-1$ non-zero components; these are $p-1$ linearly independent components of $X_i-\frac{n_i}{|n|^2}\ \vec{X}\cdot\vec{n}$.  Explicitly, for any vector $X_i$, we define
\begin{equation}
    (X_{\perp n})^r:= (T^r_{\vec{n}})^{i} X_i, \qquad \forall X_i
\end{equation} 
In this notation, the remaining constraints {become:}   
\begin{equation}\label{A-n-transverse}
\begin{split}
   K^r_{\vec{n}} := \sum_I  (K^r_{\vec{n}})^I=0,&\qquad  r=1,2,\cdots, p-1, \ \vec{n}\neq 0,\\
    (K^r_{\vec{n}})^I:= (w^I_{\perp n})^r & R_I B^I_{\vec{n}}-i \sum_{\vec{k}\neq 0,\vec{n}} \frac{1}{|\vec{k}|}({k}_{\perp n})^r A^{{I}}_{\vec{k}} B^{{I}}_{\vec{n}-\vec{k}},
\end{split}\end{equation}
where 
\begin{equation}
    (w^I_{\perp n})^r:= (T^r_{\vec{n}})^{i} w^I_i,\qquad ({{k}_{\perp n}})^r:= (T^r_{\vec{n}})^{i} k_i
\end{equation} 
$K^r_{\vec{n}}$ constitute   $p-1$ constraints which should be imposed on physical states/configurations. Notably, they have no counterpart in the null string case $p=1$. Physically, the light-cone gauge fixing employs a subset of the \( p+1 \) residual symmetries to fix the oscillator modes of the light-cone coordinates \( X^\pm \). The  $p-1$ constraints \eqref{A-n-transverse} are generators of area preserving diffeomorphisms  on the world-volume ${\cal S}_p$.

To summarize, in the light-cone gauge,  \eqref{B-n} and \eqref{A-n-parallel} specify $A^-_{\vec{n}}$ and $B^-_{\vec{n}}$. The remaining constraints include $p$ {level-matching conditions}  $L_i=0$ {from} \eqref{level-matching} and {$p-1$ transverse constraints} $K^r_{\vec{n}}=0$ {from} \eqref{A-n-transverse}, which should be imposed on physical configurations. Note that \eqref{momentum-mass} specified the light-cone mass or light-cone Hamiltonian $M^2$ and is not a constraint and that  the $K^r_{\vec{n}}=0$ may be written as vanishing of $p-1$ functions of $\sigma^i$, $K^r(\sigma^i)=0$.

\subsection{Canonical quantisation and Constraints}\label{sec:CanQuan} 

As the first step in quantisation of the null brane theory, we 
promote the  ``transverse modes'' $A^I_{\vec{n}}$ and $B^I_{\vec{n}}$ to operators ${\bf A}^I_{\vec{n}}$ and ${\bf B}^I_{\vec{n}}$, respectively, with the properties $({\bf A}^I_{\vec{n}})^\dag={\bf A}^I_{-\vec{n}}$ and $({\bf B}^I_{\vec{n}})^\dag={\bf B}^I_{-\vec{n}}$.  These operators   satisfy canonical commutation relations: 
\be
\begin{split}
[{\bf A}^I_{\vec{n}}, {\bf A}^{J}_{\vec{m}}]= 0 & = [{\bf B}^{I}_{\vec{n}}, {\bf B}^{J}_{\vec{m}}]\\
[{\bf A}^I_{\vec{n}}, {\bf B}^{J}_{\vec{m}}]= & i{|\vec{n}|}\ \delta_{\vec{m}+\vec{n},0}\delta^{IJ}.
\end{split}
\ee
Next, we introduce a new set of ``normalized oscillators'':
\begin{equation}
\bC^I_{\vec{k}}:=\sqrt{\frac{\kappa}{2 |\vec{k}|}}({\bf A}^I_{\vec{k}}+\frac{i}{\kappa} {\bf B}^I_{\vec{k}}), \qquad (\bC^I_{\vec{k}})^\dag:=\sqrt{\frac{\kappa}{2|\vec{k}|}}({\bf A}^I_{-\vec{k}}-\frac{i}{\kappa} {\bf B}^I_{-\vec{k}}),\qquad\vec{k}\neq 0.    
\end{equation}
{These oscillators obey the commutation relations:}
\begin{equation}\label{C-Cdag-commutation}
    [\bC^I_{\vec{k}}, \bC^J_{\vec{l}}]=0=[(\bC^I_{\vec{k}})^\dag, (\bC^J_{\vec{l}})^\dag], \qquad [\bC^I_{\vec{k}},(\bC^J_{\vec{l}})^\dag]=  \delta_{\vec{k},\vec{l}}\ \delta^{IJ}.
\end{equation}
Conversely, the original operators can be expressed in terms of the normalized oscillators as:
\begin{equation}
{\bf A}^I_{\vec{k}}:=\sqrt{\frac{|\vec{k}|}{2\kappa}}\left(\bC^I_{\vec{k}}+(\bC^I_{-\vec{k}})^\dagger\right), \qquad 
{\bf B}^I_{\vec{k}}:=i \sqrt{\frac{\kappa |\vec{k}|}{2}}\left((\bC^I_{-\vec{k}})^\dagger-\bC^I_{\vec{k}}\right),\qquad\vec{k}\neq 0    
\end{equation}

In terms of the ``creation'' and ``annihilation'' operators $\bC^I_{\vec{k}}$ and $\bC^J_{\vec{l}}$, we can write  the following expressions:
\begin{subequations}
\begin{align}
    \bM^2&=\sum_I\frac{m_I^2}{R_I^2}+\frac{\kappa}{2}\sum_{\vec{k}}|\vec{k}|({\mathbf{N}}_{\vec{k}}-
    \mathbf{X}_{\vec{k}}-\mathbf{X}^\dag_{\vec{k}})+ A \label{M2-C}\\
    \bL_i&= m_I w^I_i +\sum_{\vec{k}}\ k_i {\mathbf{N}}_{\vec{k}}\label{Li-C}
\end{align}
\end{subequations}
where 
\begin{equation}\label{N-X}
\mathbf{N}_{\vec{k}}=\sum_I \bC^{I\dag}_{\vec{k}}\bC^I_{\vec{k}}, \qquad \mathbf{X}_{\vec{k}}=\frac12\sum_I\bC^{I}_{\vec{k}}\bC^{I}_{-\vec{k}},\qquad 
    A= \frac{\kappa}{2} \sum_{\vec{k}} |{\vec{k}}|.
\end{equation}
{Similarly, we can} write $\bK^r_{\vec{n}}$ as: 
\begin{align}\label{K-n-C-m}
\mathbf{K}^r_{\vec{n}} &:=\sum_I (\mathbf{K}^r_{\vec{n}})^I\cr
&\, \\
(\mathbf{K}^r_{\vec{n}})^I&=i \sqrt{\frac{\kappa|\vec{n}|}{2}}  (w^I_{\perp n})^r R_I(\bC^{I\dag}_{-\vec{n}}-\bC^I_{\vec{n}})+ \frac12\sum_{\vec{k}\neq \vec{0}, \vec{n}}\sqrt{\frac{|\vec{n}-\vec{k}|}{|\vec{k}|}}\ ({k}_{\perp n})^r
\left(\bC^I_{\vec{k}}+\bC^{I\dag}_{-\vec{k}}\right)\left(\bC^{I\dag}_{\vec{k}-\vec{n}}-\bC^I_{\vec{n}-\vec{k}}\right).\nonumber
\end{align}
With these definitions, we have the following Hermiticity properties: 
\begin{equation}
(\bM^2)^\dagger=\bM^2, \;\;\;(\bL_i)^\dagger=\bL_i, \;\;\;(\bK^r_{\vec{n}})^\dagger=\bK^r_{-\vec{n}}. 
\end{equation}
\subsection{Constraints and Their Algebra}\label{sec:const-alg} 

Given the above canonical operators and their algebra, one can readily compute the algebra of operators $\bM^2, \bL_i, \bK^r_{\vec{n}}$:
\begin{subequations}\label{constraint-algebra}
\begin{align}
& [\bM^2, \bL_i]=0,\qquad [\bM^2, \bK^r_{\vec{n}}]= 0, \label{constraint-algebra-M2}\\ 
& [\bL_i, \bL_j]=0, \hspace{10mm} [\bL_i, \bK^r_{\vec{n}}]=- n_i \bK^r_{\vec{n}} \label{constraint-algebra-L}\\
&[\bK^r_{\vec{m}}, \bK^s_{\vec{n}}]=F(\vec{m},\vec{n})^{rs}_p\  \bK^p_{\vec{m}+\vec{n}}\label{constraint-algebra-K-K}
\end{align}
\end{subequations}
where $F(\vec{m},\vec{n})^{rs}_p$ is defined as
\begin{equation}
\begin{split}
    F(\vec{m},\vec{n})^{rs}_p\ (X_{\perp (m+n)})^p :=& (X_{\perp m})^r (m_{\perp n})^s -(X_{\perp n})^s (n_{\perp m})^r,\qquad \forall X_i   
\end{split}
\end{equation}
From this definition, it follows that  $F(\vec{m},\vec{n})^{rs}_p=-F(\vec{n},\vec{m})^{sr}_p$. Also, as  consistency checks, the left-hand side vanishes for $X_i=m_i+n_i$, as by definition $(X_{\perp X})^p=0$ for any $X$ and  the right-hand side also vanishes, as $((m+n)_{\perp m})^r=(n_{\perp m})^r$ and $((m+n)_{\perp n})^s=(m_{\perp n})^s$. 

We note that $\bM^2$ commutes with $\bL_i, \bK^r_{\vec{n}}$ indicating that physical states can be labelled by their $\bM^2$ eigenvalues. We will return to this point in section \ref{sec:mass}.  The operators $\bL_i$ are generator of translations on the spatial part of the brane, specifically on $T^p$ and $\bK^r_{\vec{n}}$ generate {\it {area preserving diffeomorphism}} (APD) algebra on $T^p$, SDiff($T^p$). For $p=2$, this is the APD algebra discussed in  \cite{Bars:1987dy, Bars:1988uj, deWit:1989vb}, which also admits a central extension. However, the algebra in \eqref{constraint-algebra-K-K} does not have the central extension term. 

For the later use, {consider the decomposition}  $\mathbf{K}^r_{\vec{n}}=\sum_I (\mathbf{K}^r_{\vec{n}})^I$, where { the individual components satisfy  } $[(\mathbf{K}^r_{\vec{m}})^I, (\mathbf{K}^s_{\vec{n}})^J]\propto \delta^{IJ}$. We  introduce  $p(D-2)$ mutually commuting operators  $\bL_i^I$, defined as 
\begin{equation}\label{LiI-com}
    \bL_i^{I}:=\sum_{\vec{n}} \ n_i \bC^{I\dag}_{\vec{n}}\bC^I_{\vec{n}}, \qquad [\bL_i^I, \bL_j^J]=0. 
 \end{equation}
One may note that $\bL_i=\sum_I \bL_i^{I}$ and 
   \begin{equation}\label{LiI-Krn-algebra}
    [\bL_i^I, (\bK^r_{\vec{n}})^J] =- n_i \delta^{IJ} (\bK^r_{\vec{n}})^J.  
    \end{equation}

\section{Null Brane Physical Hilbert Space}\label{sec:Hilbert-space}
After  promoting the modes ${A}^I_{\vec{n}}, {B}^I_{\vec{n}}$ and the constraints $K^r_{\vec{n}}=0=L_i$ to operators, the quantisation process is completed by constructing the physical Hilbert space. This is achieved, as usual, in three steps:  
\begin{enumerate}
    \item Define zero-excitation states (ZES); 
    \item Construct the Hilbert space by acting generic ``excitation operators'' on  ZES, to form full Hilbert space.  Denote this Hilbert space by ${\cal H}$ with a generic state represented as $|\Psi\rangle$.
    \item Define the physical Hilbert space ${\cal H}_{\text{phys}}$. Identify ${\cal H}_{\text{phys}}$ as the subset of ${\cal H}$  whose states satisfy the constraints:
\begin{equation}\label{sandwich-constraints}
\tcbset{fonttitle=\scriptsize}
            \tcboxmath[colback=white,colframe=gray]{\langle\Psi' | \bL_i|\Psi\rangle =0, \qquad \langle\Psi' | \bK^r_{\vec{n}}|\Psi\rangle =0, \qquad \forall\ |\Psi\rangle, |\Psi'\rangle \in {\cal H}_{\text{phys}}.}
\end{equation}
\end{enumerate}
It is important to emphasise that these constraints define the subset \({\cal H}_{\text{phys}} \subset {\cal H}\), rather than individual physical states. Explicitly, \textit{the constraints describe a collection of states that collectively satisfy the conditions}, as specified in \eqref{sandwich-constraints}.  

We briefly discuss in appendix \ref{appen:Sandwich-quantisation} the consistency requirements of the ``sandwich quantisation scheme,'' i.e.  imposing constraints through sandwich requirements like \eqref{sandwich-constraints}. Intriguingly, as we will discuss in detail later, this approach leads to the emergence of  various ``super-selection sectors'', corresponding to distinct ``classes'' of ${\cal H}_{\text{phys}}$ within ${\cal H}$. 

Finally, note  that, while any linear combination of physical states (i.e., generic excited states  satisfying constraints \eqref{sandwich-constraints}) is  itself  a physical state, the converse is not true: there are physical states that are linear combinations of non-physical states. In this respect, it is like having EPR entangled pairs.

\subsection{Total Hilbert Space}\label{sec:tot-Hil}

\paragraph{Zero-excitation state (ZES) $|0; m_I,w^I_i\rangle$.} These states are defined as states annihilated by all $\bC^I_{\vec{n}}, \vec{n}\neq 0$, 
\begin{equation}\label{ZES-def}
    \bC^I_{\vec{n}}|0; m_J,w^J_i\rangle=0 \qquad \forall \vec{n}\neq 0.
\end{equation}
As the notation suggests,  these states are characterised by winding and momentum labels, $m_I$ and $w^I_i$, respectively, with zero transverse excitation. However, in general one can have a ZES which is a generic linear combination of $|0; m_I,w^I_i\rangle$ with different $m_I, w^I_i$. Such states do not have a definite momentum or winding number. In this analysis, unless explicitly mentioned otherwise, we restrict ourselves to ZES with definite $m_I, w^I_i$. 

Next we note that
\be\label{Li-Krn-ZES}
\begin{split}
    \bM^2 |0; m_J,w^J_i\rangle= \big(\sum_J\ \frac{m_J^2}{R_I^2}+A\big) |0; m_J,w^J_i\rangle,&\qquad  \bL_i |0; m_J,w^J_i\rangle= \big(\sum_J\ m_Jw^J_i\big) |0; m_J,w^J_i\rangle,\\    
\bK^r_{\vec{n}}|0; m_J,w^J_i\rangle=\bigg(i \sqrt{\frac{\kappa|\vec{n}|}{2}}  (w^I_{\perp n})^r R_I \bC^{I\dag}_{-\vec{n}}+ \frac12&\sum_{\vec{k}\neq \vec{0}, \vec{n}}\sqrt{\frac{|\vec{n}-\vec{k}|}{|\vec{k}|}}\ ({k}_{\perp n})^r\bC^{I\dag}_{-\vec{k}} \bC^{I\dag}_{\vec{k}-\vec{n}}\bigg) |0; m_J,w^J_i\rangle.
\end{split}\ee
Thus, while  $|0; m_J,w^J_i\rangle$ are eigenstates of $\bM^2, \bL_i$ (and in fact eigenstates of all $\bL_i^I$), they are not eigenstates of $\bK^r_{\vec{n}}$. Therefore,  ZES are not generically  physical states, as they do not necessarily satisfy the sandwich constraints \eqref{sandwich-constraints}. We will elaborate on this in section \ref{sec:Class0ZES} for the special case of $p=2, D=4$. 

\paragraph{Generic excited states.} A generic excited state over a ZES can be written as:
\begin{equation}\label{generic-excited}
|{\bf F}; m_I, w^I_i\rangle ={\bf F}({\bC}^{J\dag}_{\vec{n}})|0; m_I, w^I_i\rangle,
\end{equation}
where ${\bf F}({\bC}^{I\dag}_{\vec{n}})$ is a generic function of all the raising operators ${\bC}^{I\dag}_{\vec{n}}$. A special subset of these states consists of ``level $N$ monomial states'' given by:
\begin{equation}\label{level-n-monomial-excitation}
|{\{{\vec{n}}_\alpha\}};m_I, w^I_i\rangle :={\bC}^{I_1\dag}_{\vec{n}_1} {\bC}^{I_2\dag}_{\vec{n}_2}\cdots {\bC}^{I_N\dag}_{\vec{n}_N}|0;{m_I,w^I_i}\rangle.
 \end{equation}

Excited states may also be constructed over a generic ZES that is itself a linear combination of $|0;m_i, w^i\rangle$ states without definite $m_I$ or $w^I_i$.  Altogether, the Hilbert space can be expressed as:
\begin{equation}\label{H-total}
{\cal H}=\bigcup_{m_I,w^I_i} {\cal H}_{m_I,w^I_i}
\end{equation}
where ${\cal H}_{m_I,w^I_i}$ is the set of all states of the form \eqref{generic-excited}. Since $\langle w^I_i, m_I;0| 0; \tilde{m}_I, \tilde{w}^I_i\rangle=\delta_{m_I,\tilde{m}_I} \delta_{w^I_i,\tilde{w}^I_i}$ all states in ${\cal H}_{m_I,w^I_i}$ and ${\cal H}_{\tilde{m}_I,\tilde{w}^I_i}$ with different $m, w$'s are orthogonal to each other. 

As mentioned not all states in ${\cal H}$ \eqref{H-total} are physical. Let us denote the set of all physical states in the Hilbert space ${\cal H}$ by ${\cal H}_{\text{phys}}$ and its complement by ${\cal H}^c_{\text{phys}}$, i.e.  ${\cal H}={\cal H}_{\text{phys}} \cup {\cal H}^c_{\text{phys}}$. One may argue that, without loss of generality, all states in ${\cal H}_{\text{phys}}$ and ${\cal H}^c_{\text{phys}}$ can be consistently taken  to be orthogonal, i.e.
\begin{equation}\label{HHc-orthogonal}
    \langle\Phi|\Psi^c\rangle=0\qquad \forall\ |\Phi\rangle\in {\cal H}_{\text{phys}} , \ |\Psi^c\rangle\in {\cal H}^c_{\text{phys}}.
\end{equation}

Finally, a generic state in ${\cal H}$, upon imposing the constraints \eqref{sandwich-constraints}, can be interpreted as a ``single null brane'' state. However,  multi-brane states can also be considered within the null brane Fock space, ${\cal F} = \bigcup_{n \in \mathbb{Z}^*} {\cal H}^n$, where $\mathbb{Z}^*$ denotes non-negative integers.  
Physical Fock space, ${\cal F}_{\text{phys}}$, is then the union of physical subspaces ${\cal H}_{\text{phys}}^n$. It is immediate to see that a generic (physical or not) $m$ brane state 
and $n$-brane state are orthogonal {for} $m\neq n$, a feature of free (non-interacting) theories.   Interaction operators yield a non-zero overlap between generic multi brane states. Since in this {work} we  only {focus} on the free brane theory, we restrict attention to   single-brane states and study {the} physicality conditions for states in ${\cal H}$.

\subsection{Imposing Sandwich Constraints, a General Setup}\label{sec:Class0-general}

Consider a generic physical observable operator $\boldsymbol{{\mathcal{O}}}$ acting on a state in  ${\cal H}_{\text{phys}}$. The action of $\boldsymbol{{\mathcal{O}}}$ generally yields a state in ${\cal H}_{\text{phys}}$ plus a state in ${\cal H}^c_{\text{phys}}$, expressed as
\begin{equation}\label{O-1}
    \boldsymbol{{\mathcal{O}}} |\Psi\rangle = |\Psi_{\cal O}\rangle  +|\Psi_{\cal O}^c\rangle,  \qquad |\Psi\rangle, |\Psi_{\cal O}\rangle \in {\cal H}_{\text{phys}},\quad |\Psi^c_{\cal O}\rangle\in {\cal H}^c_{\text{phys}}.
\end{equation}
We note that a generic observable $\boldsymbol{{\mathcal{O}}}$ is constructed from the operators $\bC^I_{\vec{k}}, (\bC^I_{\vec{k}})^\dag$. Therefore, the states $|\Psi\rangle,|\Psi_{\cal O}\rangle$ and $|\Psi_{\cal O}^c\rangle$ all belong to the same subspace ${\cal H}_{m_I,w^I_i}$ (characterised by the same ${m_I,w^I_i}$).  

The physicality condition of states is defined through vanishing of {the} constraint operators $\bL_i$ and $\bK^r_{\vec{n}}$, sandwiched between any two physical states \eqref{sandwich-constraints}. Thus, as noted from \eqref{HHc-orthogonal} and \eqref{O-1}, the action of $\bL_i$ or $ \bK^r_{\vec{n}}$  on a physical state does not necessarily yield another physical state. Instead, there may be a component $|\Psi^c\rangle$ that lies in ${\cal H}^c_{\text{phys}}$, which does not contribute to the physicality condition.

To proceed, observe that a generic state in the Hilbert space can be labelled by eigenvalues of $\bL_i$. In what follows we use ``zero eigenstates'' as a shorthand for eigenstates with vanishing eigenvalues. So, one can in general classify the states into $p+1$ classes:
\begin{itemize}
    \item \textbf{Class 0} \textit{Consisting of states that are zero eigenstates of all $\bL_i$.}
\item  \textbf{Class} $\bcN$, $\bcN=1,2,\cdots, p$ \textit{Consisting of states with $\bcN$ number of the $\bL_i$ have non-vanishing eigenstates, i.e. $p-\bcN$  of the $\bL_i$ have vanishing eigenstates.}
\end{itemize}
However,  not all states within these classes are physical.  Whether a state satisfies the physicality conditions must be verified against the constraints in \eqref{sandwich-constraints}.

\paragraph{Proposition 1.}    \textit{\textbf{Class 0} states, defined as the set of all zero-eigenstates of $\bL_i$,  form a physical Hilbert space.}

\paragraph{Proof.} Consider two generic zero-eigenstates of $\bL_i$ and $|\Psi\rangle, |\Phi\rangle$, i.e.  $\bL_i|\Psi\rangle=\bL_i|\Phi\rangle=0$. Then, from \eqref{constraint-algebra-L} we learn that,
\begin{equation}
    \langle \Phi|[\bL_i, \bK^r_{\vec{n}}] |\Psi\rangle=0=-n_i  \langle \Phi|\bK^r_{\vec{n}}|\Psi\rangle. 
\end{equation}
Since $n_i \neq 0$ for $\bK^r_{\vec{n}}$ the constraints  \eqref{sandwich-constraints} are satisfied for the set of zero-eigenstates of $\bL_i$. 

We crucially note that these physical states need \emph{not} be, and generically do not, satisfy the constraints as  eigenstates of $\bK_{\vec{n}}$. Specifically, while \textbf{Class 0} states  satisfy $\bL_i$ constraints through the ``right-action'' condition $\bL_i|\Psi\rangle=0, \ \forall |\Psi\rangle \in {\cal H}_{\text{phys}}$, the $\bK_{\vec{n}}$ constraints are not fulfilled through the right-action, they are satisfied only as the sandwich conditions $\langle\Phi|\bK_{\vec{n}}|\Psi\rangle=0$.

\subsection{Beyond Class \texorpdfstring{$\bcN= 0$}{}}\label{sec:Class-N}

When $\bL_i|\Psi\rangle \neq 0$ for some of the $\bL_i$, the $ \bL_i$ sandwich constraints can be solved using the methods described in \cite{Bagchi:2024tyq}. However, the $\bK^r_{\vec{n}}$ sandwich constraints are  significantly  more involved, and the ideas from \cite{Bagchi:2024tyq} are not applicable. In the following,  we develop new techniques for dealing with $\bK^r_{\vec{n}}$ sandwich constraints. {Our approach begins by solving} the $\bL_i$ sandwich constraints, after which we address the $\bK^r_{\vec{n}}$ constraints. 

To analyse the problem further and implement the method from \cite{Bagchi:2024tyq}, recall that $\bL_i$ are Hermitian operators on the total Hilbert space ${\cal H}$ and one can use their eigenstates as a complete basis over ${\cal H}$. Explicitly, 
\begin{equation}\label{Li-eigenstates1}
    \bL_i|\ell_i\rangle= \ell_i |\ell_i\rangle,\qquad  i=1,2,\cdots, p.
\end{equation}
Note that for a given eigenvalue  $\ell_i$, the eigenstates $|\ell_i\rangle$ are highly degenerate. For example, in the case of monomial states \eqref{level-n-monomial-excitation}, the eigenvalue is given by $\ell_i= m_I w^I_i+ \sum_{M} (n_M)_i$ and this value is insensitive to the specific  details of the indices $I_1, I_2, \cdots$  of the state. 

We can partially resolve this degeneracy  by considering the eigenstates of mutually commuting operators $\bL_i^I$ \eqref{LiI-com}:
\begin{equation}\label{LIi-eigenstates}
    \bL^I_i|\ell^I_i\rangle= \ell^I_i |\ell^I_i\rangle,\qquad i=1,2,\cdots, p, \quad I=1,2,\cdots D-2.
\end{equation}
These eigenstates can  be rearranged into eigenstates corresponding to linearly independent sets  of operators $\bL_i$ and $\bL_i^A$ where $A=1,2,\cdots, D-3$. So, we may denote these eigenstates as $|\ell^I_i\rangle=|\ell_i; \ell_i^A\rangle$, with the eigenvalues satisfying 
\begin{equation}\label{Li-LiA}
\ell_i=\sum_I \ell^I_i,\qquad \ell_i^A=  \text{Combinations of}\ \ell_i^I\  \text{linearly independent of \st{with} } \ell_i.
\end{equation}
In this notation, a generic state in ${\cal H}$ may be denoted as
\begin{equation}
|\Psi\rangle = 
\sum_{\ell_i, \ell_i^A}\ \Psi(\ell_i; \ell_i^A) |\ell_i; \ell_i^A\rangle
\end{equation}
It is important to note that while $\ell_i$ has a fixed/given form in terms of $\ell_i^I$, as given by \eqref{Li-LiA}, the components $\ell_i^A$ are not uniquely fixed in terms of $\ell_i^I$.  While this ambiguity/freedom is not of any significance at this stage, it will have implications for the physical states we construct in the following sections. We will explore this point in detail in section \ref{sec:counting-revisited}, where we focus on the specific case $p = 2, D = 4$.

\subsubsection{Class \texorpdfstring{$p$}{} States}\label{sec:Class-p}

In the \textbf{Class} $\bcN=p$ case, we are dealing with $\ell_i\neq 0$ for $ i=1,2,\cdots, p$ states where $\ell_i$ take both positive and negative values. Let  the sign of $\ell_i$ be denoted by $s_i$, i.e. $\ell_i=|\ell_i| s_i$. One may define generic \textit{standing waves} as follows\footnote{Recall that $\bL_i$ are generator of translations on the (spatial part of the) worldvolume and hence $\ell_i$ are basically denote momenta on the worldvolume of the brane.} 
\begin{equation}\label{standing wave -Li}\begin{split}
    ||\ell_i|; \ell_i^A \rangle_{_{\text{\tiny{S}}}}:=\frac{1}{2^{p-1}}\ \sum_{s_i=+, -} ||\ell_i|s_i;  \ell_i^A\rangle, \qquad |\ell_i|\geq 0.
\end{split}
\end{equation}
There are $2^p$ terms in the sum, as there are $p$ signs $s_i$,  each {of which can} take values $+$ or $-$.  A generic standing wave state may be written as 
\begin{equation}\label{standing wave -Li1}\begin{split}
    |\psi(|\ell_i|; \ell_i^A) \rangle_{_{\text{\tiny{S}}}}:=\sum_{\ell_i, \ell_i^A}\psi(|\ell_i|; \ell_i^A)\  ||\ell_i|; \ell_i^A \rangle_{_{\text{\tiny{S}}}}.
\end{split}
\end{equation}

Standing waves \eqref{standing wave -Li} are not eigenstate of $\bL_i$, although they can be eigenstates of $\bL_i^A$. Specifically, for any  given $j$, there are $2^{p-1}$ terms like $||\ell_i|s_i;  \ell_i^A\rangle$ with positive $\bL_j$ eigenvalue $|\ell_j|$ and $2^{p-1}$ terms with negative eigenvalue $-|\ell_j|$.
While standing waves  are not eigenstates of $\bL_i$, they are eigenstates of $(\bL_i)^2$. These properties are summarised in the following equations. 
\begin{subequations}
\begin{align}
\bL_j  ||\ell_i|; \ell_i^A \rangle_{_{\text{\tiny{S}}}} & =\frac{|\ell_j|}{2^{p-1}}\ \sum_{s_i=+, -}\  s_i\ ||\ell_i|s_i;  \ell_i^A\rangle, \label{Lj-standingwave}\\ 
\bL_j^B  ||\ell_i|; \ell_i^A \rangle_{_{\text{\tiny{S}}}} & = \ell_j^B\ ||\ell_i|; \ell_i^A \rangle_{_{\text{\tiny{S}}}}\\
(\bL_j)^2\ ||\ell_i|; \ell_i^A  \rangle_{_{\text{\tiny{S}}}} &= (\ell_j)^2\ ||\ell_i|; \ell_i^A\rangle_{_{\text{\tiny{S}}}},\qquad \text{no sum over} \; j.
    \end{align}
\end{subequations} 
From  \eqref{Lj-standingwave}, it follows that: 
\begin{equation}
  {}_{_{\text{\tiny{S}}}}\langle \tilde{\psi}(|\tilde\ell_i|; \ell_i^A) \ |\ \bL_j\ |\psi(|\ell_i|; \ell_i^A )\rangle_{_{\text{\tiny{S}}}}=0,\qquad \forall  \psi, \tilde{\psi}.
\end{equation}
That is, the standing waves states $|\psi(|\ell_i|;  \ell_i^A)\rangle_{_{\text{\tiny{S}}}}$ are constructed to satisfy the $\bL_i$ sandwich conditions \eqref{sandwich-constraints}.

\paragraph{Tackling  $\bK^r_{\vec{n}}$ sandwich constraints.}  We explore the idea that the  $\bK^r_{\vec{n}}$ constraints require fixing certain $\ell_i^A$ parameters from the standing wave solutions that solve the \(\bL_i\) sandwich constraints.  Explicitly, we require that these states are eigenstates of a subset of $\ell_i^A$, henceforth labelled by $\ell^a$, $a=1,...,m$, 
\begin{equation}\label{semifinal-class-II}
    |\ell^a; \psi(|\ell_i|;\ell^\alpha)\rangle_{_{\text{\tiny{S}}}} :=\sum_{\ell^\alpha}\psi(|\ell_i|; \ell^\alpha)\ ||\ell_i|;\ell^a; \ell^\alpha\rangle_{_{\text{\tiny{S}}}},
\end{equation}
such that 
\begin{equation}\label{li-la-l-alpha}
    \bL^a |\ell^a; \psi(|\ell_i|;\ell^\alpha) \rangle_{_{\text{\tiny{S}}}}=\ell^a |\ell^a;  \psi(|\ell_i|; \ell^\alpha)\rangle_{_{\text{\tiny{S}}}}.
\end{equation}
Here, $\ell^\alpha$ denotes the remaining part of $\ell_i^A$'s after fixing $\ell^a$, so $\alpha=1,2,\cdots, p(D-3)-m$.

Our analysis utilises the  commutation relations in \eqref{LiI-Krn-algebra}. To facilitate the argument, we decomposed $\bL_i^I$ into $\bL_i$ and $\bL_i^A$. $(\bK^r_{\vec{n}})^I$ may be decomposed into $\bK^r_{\vec{n}}=\sum_I (\bK^r_{\vec{n}})^I$ and the remaining $D-3$ linearly independent parts into $(\bK^r_{\vec{n}})^A$. In this notation, \eqref{LiI-Krn-algebra} may be decomposed into
\begin{subequations}\label{LiA-KrnI-algebra}
\begin{align}
[\bL_i, \bK^r_{\vec{n}}] =- n_i   \bK^r_{\vec{n}},& \qquad  [\bL_i, (\bK^r_{\vec{n}})^A]  =- n_i   (\bK^r_{\vec{n}})^A, \label{Li-KrnI-algebra}\\
     [\bL_i^A, \bK^r_{\vec{n}}] &=-  n_i (\bK^r_{\vec{n}})^A,\label{LiA-Krn-algebra}\\
  [\bL_i^A, (\bK^s_{\vec{n}})^B]  &=- n_i  \delta^{AB} (\bK^s_{\vec{n}})^B\ . \label{LiA-KrnB-algebra}
    \end{align}
\end{subequations}
Since standing waves are not eigenstates of $\bL_i$, in our arguments below we will use \eqref{LiA-Krn-algebra} and \eqref{LiA-KrnB-algebra}.

The key question is then, which $\ell^a$ combinations in \eqref{li-la-l-alpha} should be fixed to satisfy the $\bK_{\vec{n}}^r$ constraints. To address this,  we take $\ell^a$'s as linear combinations:
\begin{equation}\label{TaiA}
    \ell^a=\sum_{A,i}\ (T^{a})^i_A\ \ell^A_i.
\end{equation}
and try to find $(T^{a})^i_A$. Sandwiching \eqref{LiA-Krn-algebra} and \eqref{LiA-KrnB-algebra} between two standing wave solutions yields:
\begin{equation}
\begin{split}
  \big(\tilde\ell^a-\ell^a\big)\ \langle \bK_{\vec{n}}^r\rangle  & =-\big(\sum_j n_j (T^{a})^j_A \big)\ 
  \langle (\bK_{\vec{n}}^r)^A\rangle, \\
   \big(\tilde\ell^a-\ell^a\big)\ \langle (\bK_{\vec{n}}^r)^B\rangle 
  & =-\big(\sum_j n_j (T^{a})^j_B \big)\ \langle (\bK_{\vec{n}}^r)^B\rangle,\qquad \text{no sum over } B
\end{split}
\end{equation}
where
\begin{equation}\begin{split}
   \langle \bK_{\vec{n}}^r\rangle &:=  {}_{_{\text{\tiny{S}}}}\langle \tilde\psi(|\tilde\ell_i|; \ell^\alpha);  \tilde\ell^a |\  \bK^r_{\vec{n}}\  | \ell^a; \psi(|\ell_i|; \ell^\alpha) \rangle_{_{\text{\tiny{S}}}},\\
  \langle (\bK_{\vec{n}}^r)^A\rangle &:=  {}_{_{\text{\tiny{S}}}}\langle \tilde\psi(|\ell_i|; \ell^\alpha); \tilde\ell^a|\  (\bK^r_{\vec{n}})^A\  |  \ell^a; \psi(|\ell_i|;\ell^\alpha) \rangle_{_{\text{\tiny{S}}}}\,. 
\end{split}\end{equation}
For the constraints to hold  for all $n_i$, the two sides should be zero independently, i.e. $\tilde\ell^a=\ell^a$  and 
\begin{equation}
    (\sum_i n_i (T^{a})^i_A)\langle (\bK_{\vec{n}}^r)^A\rangle \ =0, \qquad \forall n_i, A, a.
\end{equation}
For fixed $A$, $\sum_i n_i(T^{a})^i_A$ is a linear combination of $n_i$, $i=1,...,p$. Therefore if we have $p$ linearly independent combinations, i.e. if $m=p$ ($\# a=p $), for any $\vec{n}\neq \vec{0}$ at least one of these for some $a$ will be non-zero and the $\bK^r_{\vec{n}}$ constraints will be fulfilled. 

Finally, we recall discussions of appendix \ref{appen:Sandwich-quantisation} and in particular \eqref{Operator-Product}. We need to make sure that product of two operators creating the standing waves in \eqref{semifinal-class-II} acting on the ZES is another state of the same form. This can't be satisfied unless all $\ell^a$ are set to zero, that is \textbf{Class} $\bcN=p$ are states of the form,
\begin{equation}\label{final-class-II}
    |0; \psi(|\ell_i|;\ell^\alpha)\rangle_{_{\text{\tiny{S}}}} :=\sum_{|\ell_i|,\ell^\alpha}\psi(|\ell_i|; \ell^\alpha)\ ||\ell_i|;0; \ell^\alpha\rangle_{_{\text{\tiny{S}}}}.
\end{equation}

\subsubsection{Class \texorpdfstring{$\bcN\neq 0, p$}{} States}\label{sec:Class-N-not0p}

It may {happen}  that some of the $\ell_i$ in the standing wave part are vanishing. Recalling the commutation relations in \eqref{Li-KrnI-algebra}, the above argument should be revisited for such cases. Suppose  $\ell_i\neq 0$ for $1\leq i\leq \bcN$ and  $\ell_i= 0$ for $\bcN+1\leq i\leq p$. Sandwiching \eqref{Li-KrnI-algebra} between standing waves  with $p-\bcN$ vanishing $\ell_i$, for $n_{1}\neq 0$ or ... $n_{p-\bcN}\neq 0$ the $\bK^r_{\vec{n}}$ constraint is automatically satisfied due to $\ell_{1}=\cdots=\ell_{p-\bcN}=0$. Therefore, the only non-trivial cases to study {occur when} $n_{\bcN+1}=\cdots=n_p=0$,  $\bcN$ number of the $(T^{a})_i^A$ coefficients are required to be linearly independent. Thus, {the} number of $\ell^a$ that should be fixed to zero in this sector is $\bcN$. Given that in this sector $p-\bcN$ number of {the} $\ell_i$ are already fixed (to zero), there  {remain}  $p$ number of $\ell_i^I$ which {are} vanishing in general. 

Explicitly, in the sector with $p-\bcN$ vanishing $\ell_i$, {the} physical states are of the form
\begin{equation}\label{final-class-II-q}
            {    |\bcN; \psi(|\ell_\natural|;\ell^\flat)\rangle_{_{\text{\tiny{S}}}} :=\sum_{|\ell_\natural|,\ell^\flat}\psi(|\ell_\natural|;\ell^\flat)\ ||\ell_\natural|;\ell^\flat\rangle_{_{\text{\tiny{S}}}},\quad \natural =1,\cdots, \bcN,\ \ \flat=1,2,\cdots, \ p(D-3)-\bcN}
\end{equation}
where $||\ell_\natural|;\ell^\flat\rangle_{_{\text{\tiny{S}}}}$ denotes a standing wave  like \eqref{standing wave -Li} with $2^{\bcN}$ terms in the sum with alternating signs on $\ell_\natural$ values, and with $\ell^\flat$ (among the $\ell^A_i$) taking arbitrary non-zero values. 

\paragraph{Summary of all physical \textbf{Class} $\bcN$, $\bcN=0,1,\cdots, p$ states.} As {argued} above, one can readily identify  \textbf{Class 0} states as \textbf{Class} $\bcN=0$.  Therefore, \textit{all} physical states can be expressed uniformly as   
\begin{equation}\label{final-class-I-class-II}
\tcbset{fonttitle=\scriptsize}
            \tcboxmath[colback=white,colframe=gray]{    |\bcN; \psi(|\ell_\natural|;\ell^\flat)\rangle_{_{\text{\tiny{S}}}} :=\sum_{|\ell_\natural|,\ell^\flat}\psi(|\ell_\natural|;\ell^\flat)\ ||\ell_\natural|;\ell^\flat\rangle_{_{\text{\tiny{S}}}},\quad \natural =1,\cdots, \bcN,\ \ \flat=1,2,\cdots, \ p(D-3)-\bcN}
\end{equation}
with $\bcN=0,1,\cdots, p$. In each $\bcN$-sector a physical state is labelled by $p(D-3)$ integers, $\bcN$ positive integers and $p(D-3)-\bcN$ integers of arbitrary sign. The positive integers may be selected from  $\ell_i$ (there are $p$ of them) while the remaining $p(D-3)-\bcN$ integers are chosen  among $\ell^I_i$ which are linearly independent of the $\ell_i$. 

We close by the comment that  we argued for the existence of $p-\bcN$ number of $(T^a)_A^i$ matrices and did not explicitly identify them. The number of super-selection sectors depends not only on the possible choices of $\bcN$ from $p$ (or $\bcN$ from $p(D-3)$) but also on the number of independent $(T^a)_A^i$ matrices for a given $\bcN$.  Moreover, in our case we have considered toroidal $p$-branes on a toroidally compact target space and thus  $\ell^I_i$ take integer values and $\ell_i^I$ are defined modulo $SL(p,\mathbb{Z})$ transformations on the on the $i$ index.   We will present an explicit counting of super-selection sectors  for the case of $p=2, D=4$, taking these comments into account, in section \ref{sec:counting-revisited}.

\subsection{Mass of Physical States}\label{sec:mass} 

So far, we have classified all physical states as described  in \eqref{final-class-I-class-II}.  These states are organised into \textbf{Class} $\bcN$ sectors labelled by $\bcN=0,1,\cdots, p$, , where $p-\bcN$ being the number of $\bL_i$ with zero eigenstates. Within each sector, states are further specified by $p(D-3)$ integers.  Here we discuss the assignment of mass to physical null $p$-brane states. Regarding the mass of null branes, a few observations are in order:  
\begin{enumerate}
    \item  As reviewed in section \ref{sec:null-brane-action}, null $p$-branes may be understood as a certain tensionless limit of tensile $p$-branes. While null branes are tensionless, not all tensionless branes are null. In particular, note that the null $p$-brane action \eqref{null-brane-action-covariant} has a dimensionful parameter  $\kappa$ {with} dimension of mass$^{2}$.
    \item As discussed in section \ref{sec:p-brane-EoM-null-congruence}, the worldvolume of a null $p$-brane  consists of a congruence of null geodesics in the $D$-dimensional target space. From this perspective,  one  {might} expect null $p$-branes to be massless from target space viewpoint. However, as in the {standard} massless particle cases, one can define light-cone Hamiltonian/energy, which serves  as the effective mass of the state. Indeed, what we have called $M^2$ is precisely this light-cone Hamiltonian. This is analogous to the mass assignment for string states in the light-cone gauge \cite{Green:1987sp, Polchinski:1998rq}.  
\end{enumerate} 

One may use operator $\mathbf{M}^2$ to define the mass of physical states. As already noted, while in \eqref{constraint-algebra} we have included commutators of $\mathbf{M}^2$ with $\bL_i, \bK^r_{\vec{n}}$,  $\mathbf{M}^2$ is not among our constraints. Moreover,  one should be take a special care in defining the mass for null $p$-brane states. Below we outline  two possible definitions and justify  adopting  the second possibility.

\paragraph{Mass as eigenvalue of $\mathbf{M}^2$.} {The} commutators in \eqref{constraint-algebra-M2} imply that $\bM^2$ and $\bL_i$ can be diagonalised simultaneously. Furthermore, 
\begin{equation}\label{M2-LiI}
    [\bM^2, \bL_i^I]=0, 
\end{equation}
indicating that physical states can consistently be eigenstates of $\bM^2$. Consequently, a label $M$ can be added to the states, denoted as $|M; |\ell_\natural|; \ell^\flat\rangle_{_{\text{\tiny{S}}}}$. 
We note that since $[\bM^2, \bK^r_{\vec{n}}]=0$, eigenstates of $\bM^2$ are compatible with the $\bK^r_{\vec{n}}$ sandwich conditions. That is, satisfying the $\bK^r_{\vec{n}}$ sandwich conditions requires selecting states within a fixed $M$ sector. However, recalling \eqref{M2-C} and \eqref{N-X}, note that  $\mathbf{X}_{\vec{k}}$ contains two annihilation  operators. Therefore, states with a given excitation number $\mathbf{N}_{\vec{k}}$  are not strictly  eigenstates of $\bM^2$; instead, eigenstates of $\bM^2$ resemble squeezed states \cite{Bagchi:2020fpr, Bagchi:2020ats}. 

\paragraph{Mass as expectation value of $\mathbf{M}^2$.}  
We remark that \eqref{M2-C} is a result of {the} light-cone gauge fixing,  which allows solving for $B^0_-$ and obtaining $M^2$ \eqref{momentum-mass}. Upon quantisation, this conditions, can also be 
viewed as fulfilled through sandwich conditions or the ``right action'' (see appendix \ref{appen:Sandwich-quantisation}). 

This suggests an alternative way to associate mass with physical states, in line with the  sandwich conditions and consistent with the sandwich constraint quantisation. Specifically, the mass of a normalised physical state $|\Psi\rangle$ can be \textit{defined} as 
\begin{equation}\label{mass-def}
\tcbset{fonttitle=\scriptsize}
            \tcboxmath[colback=white,colframe=gray]{   
            M^2_\Psi:= \langle \Psi|\mathbf{M}^2 |\Psi\rangle}  
\end{equation}

States labelled by $p(D-3)$ and $\ell_i^I$ exhibit high degeneracy, with many states sharing the same mass (as defined in \eqref{mass-def}) and $\ell_i^I$ labels. Nonetheless, the mass of states at a given excitation level $N$ can still be computed. Suppose 
\begin{equation}
    (\sum_{\vec{k}} |\vec{k}| {\mathbf{N}}_{\vec{k}}) |\Psi\rangle = N_\Psi |\Psi\rangle
\end{equation}
where ${\mathbf{N}}_{\vec{k}}$ is defined in \eqref{N-X}. Moreover, the commutators  
\begin{equation}
    [{\mathbf{N}}_{\vec{k}}, {\mathbf{X}}_{\vec{p}}]=-{\mathbf{X}}_{\vec{k}} \left(\delta_{\vec{k},\vec{p}}+\delta_{\vec{k},-\vec{p}}\right),\qquad [{\mathbf{X}}_{\vec{k}}, {\mathbf{X}}^\dagger_{\vec{p}}]=\frac14 ({\mathbf{N}}_{\vec{k}}+{\mathbf{N}}_{\vec{-k}}) \left(\delta_{\vec{k},\vec{p}}+\delta_{\vec{k},-\vec{p}}\right)
\end{equation}
imply that for eigenstates of ${\mathbf{N}}_{\vec{k}}$
\begin{equation}\label{mass-psi}
    M^2_\Psi:= \sum_I \frac{m_I^2}{R_I^2}+ \frac{\kappa}{2} N_\Psi - \frac{\kappa}{2}\sum_{\vec{k}} |k|\  \langle \Psi|({\mathbf{X}}_{\vec{k}}+{\mathbf{X}}^\dagger_{\vec{k}}) |\Psi\rangle= \sum_I \frac{m_I^2}{R_I^2}+ \frac{\kappa}{2}N_\Psi.
\end{equation}
As remarked earlier, the null $p$-brane tension $\kappa$ appears {explicitly} in the expression for $M^2$. 

\section{The Example of Null Membranes in \texorpdfstring{$D=4$}{} }\label{sec:p2D4}
In the preceding section, we presented a formal solution to the sandwich constraints \eqref{sandwich-constraints}  and categorized the solutions into $p+1$ \textbf{Class} $\bcN$ states for a generic null p-brane in $D$ dimensions. 
 
To make the formal analysis more tractable in this section we  examine in greater detail the {specific} case of a $p=2$ membrane in $D=4$. We begin by constructing the total Hilbert space and then proceed to explicitly classify the \textbf{Class 0}, \textbf{Class 1} and \textbf{Class 2} states. In this case, both $I$ and $i$ indices run over the same values $1, 2$. There is only one $\bK_{\vec{n}}$ generator and the constraint algebra takes the following form: 
\begin{equation}\label{constraint-algebra-2d}
\begin{split}
& [\bM^2, \bL_i]=0,\qquad [\bM^2, \bK_{\vec{n}}]= 0,\\ 
& [\bL_i, \bL_j]=0, \hspace{10mm} [\bL_i, \bK_{\vec{n}}]=- n_i \bK_{\vec{n}},\\
& [\bK_{\vec{m}}, \bK_{\vec{n}}]=-i \epsilon_{ij} m_i n_j \bK_{\vec{m}+\vec{n}},
\end{split}
\end{equation}
where $\epsilon^{ij}$ is the $2d$ antisymmetric tensor. 

We note that $\bM^2$ commutes with {both} $\bL_i$ and $\bK_{\vec{n}}$ indicating that physical states can be labelled by their mass. The {operators} $\bL_i$ generate translations on the spatial part of the membrane, the $T^2$, while  $\bK_{\vec{n}}$ generate the {algebra of} {\it {area preserving diffeomorphism}}  on $T^2$, SDiff($T^2$) \cite{Bars:1987dy, Bars:1988uj, deWit:1989vb} which admits a central extension. 

\subsection{\textbf{Class 0} Hilbert Space of Physical States}

In section \ref{sec:Class0-general}, we discussed that \textbf{Class 0} consists of  generic states with $\ell_1, \ell_2=0$. Here we explicitly construct these states for a null membrane in $D=4$. We start with the zero-excitation (ZES) and then those with generic excitation numbers.

\subsubsection{Physical Zero-Excitation States}\label{sec:Class0ZES}

Zero-excitation states (ZESs), denoted by $|0; m_I, w_j^I\rangle$, are labelled by different values of $m_I$ and  $w^I_j$. There states are eigenstates of $\bL_i$ but not of $\bK_{\vec{n}}$ {or} $\bM^2$. {While} $|0; m_I, w_j^I\rangle$ states are not {necessarily} physical, {those satisfying physicality conditions (sandwich constraints)  belong to one of \textbf{Class 0}, \textbf{Class 1}, or \textbf{Class 2}. 
A ZES {belongs to} \textbf{Class 0} if it satisfies the condition   $\sum_{I=1,2} m_I w_j^I=m_1 w^1_j+m_2 w^2_j=0$ for $j=1,2$. These two equations have three classes of solutions, zero-momentum solution $m_1=m_2=0$,  one of $m_1$ or $m_2$ is zero and none {$m_1$ or $m_2$} is zero, in which case one can be solved  in terms of the other.  
\begin{enumerate}
    \item Zero-momentum vacua $|0; m_I=0, w_j^I\rangle$, with $w_j^I$ arbitrary integers; 
    \item Mixed vacua with one of $m_1$ or $m_2$ is zero. If $m_1=0, m_2\neq 0$, then $w^2_j=0$ with $w^1_j$ arbitrary. And similarly for the $m_2=0, m_1\neq 0$ case.
    \item Generic vacua when $m_1$ and $m_2$ are both non-zero and $w^1_j=-\frac{m_2}{m_1} w^2_j$.
\end{enumerate} 
{None} of the above physical ZESs are eigenstates of the mass operator  $\bM^2$. Nonetheless, employing 
\eqref{mass-def} and \eqref{mass-psi} one can associate a mass with {any of} these ZESs: {The} mass of {a state} $|0; m_I, w_j^I\rangle$ is {given by} $M^2=\sum_I m_I^2/R_I^2$. Notably, the mass of ZESs depends only on $m_I$ and not on $w^I_j$. Therefore, the zero-momentum states $|0; m_I=0, w^I_j\rangle$ all have degenerate, vanishing mass and are  the lowest-mass physical states.  

\subsubsection{Excited \textbf{Class 0} States}\label{sec:Class0-generic-excitation}

We explicitly construct excited \textbf{Class 0} states through the following two propositions.

\paragraph{Proposition 3.} \textit{Let $|\Psi\rangle$ be a generic \textbf{Class 0} physical state,  then ${\bf F}({\bC}^{I\dag}_{\vec{p}}) |\Psi\rangle$ is also a \textbf{Class 0} physical state iff  $[\bL_i, {\bf F}({\bC}^{I\dag}_{\vec{p}})]=0$.}

\paragraph{Proof.}  In order ${\bf F}({\bC}^{I\dag}_{\vec{p}}) |\Psi\rangle$ to qualify as a \textbf{Class 0} physical state, it should be a zero eigenstate of $\bL_i$, as per Proposition 1. Since  $\bL_i |\Psi\rangle=0$, the condition is satisfied if $\bL_i {\bf F}({\bC}^{I\dag}_{\vec{p}})= \tilde{{\bf F}}({\bC}^{I\dag}_{\vec{p}})\bL_i$, ensuring that $\bL_i{\bf F}({\bC}^{I\dag}_{\vec{p}}) |\Psi\rangle=0$. However, recalling the form of $\bL_i$ \eqref{Li-C}, we find that $\tilde{{\bf F}}({\bC}^{I\dag}_{\vec{p}})$ must be ${\bf F}$. To this end we note that \eqref{canonical-Poisson-brackets} implies $\bC^I_{\vec{k}}=\frac{\partial}{\partial (\bC^I_{\vec{k}})^\dag}$ and hence 
\begin{equation}
    [\bL_i, {\bf F}({\bC}^{I\dag}_{\vec{p}})]= \sum_{\vec{k}} k_i ({\bC}^{I\dag}_{\vec{k}}) \frac{\partial  {\bf F}}{ \partial({\bC}^{I\dag}_{\vec{k}})}
\end{equation}
and the RHS of the above is only a function ${\bC}^{I\dag}_{\vec{p}}$ which does not vanish acting on any physical state, unless the RHS is identically zero. 

In the next proposition we construct all possible functions ${\bf F}$ that commute with $\bL_i$.

\paragraph{Proposition 4.} \textit{Any operator that commutes with $\bL_i$ is a function of the set of operators $\mathbf{Y}_n$'s, defined as} 
\begin{equation}\label{Yn}
    \mathbf{Y}_n(\{\vec{p}_a, j_a\}):=\left(\prod_{a=1}^{n} ({C}^{J_a}_{\vec{p}_a})^\dag\right), \qquad \sum_{a=1}^n p^{j{_{_a}}}=0,\quad \text{for}\ j=1,2, \quad \text{and}\ \ n\geq 2.
\end{equation}
\textit{That is,} $[\bL_i, \mathbf{F}(\{\mathbf{Y}_n (\{\vec{p}_a, j_a\})\} )]=0$.

\textbf{Proof.} We begin by demonstrating that $[\bL_i, \mathbf{Y}_n]=0$. To do so, consider the following computation:
\begin{equation}\begin{split}
    [\bL_i, \mathbf{Y}_n(\{\vec{p}_a, j_a\})] &=\sum_{\vec{k}}\ k_i (\bC^I_{\vec{k}})^\dag \frac{\partial}{\partial (\bC^I_{\vec{k}})^\dag}\ \mathbf{Y}_n \\ &=    \left(\sum_{a=1}^n\ (p^{j_a}) \right)\ \mathbf{Y}_n(\{\vec{p}_a, j_a\})=0,
\end{split}
\end{equation}
where we used the fact that $\sum_{a=1}^n p^{j_a}=0$ in the last equality. It is important to note that  $p^{j_a}\neq 0$ and that $n\geq 2$. Next, we observe that
\begin{equation}
    [\bL_i, \mathbf{F}(\{\mathbf{Y}_n (\{\vec{p}_a, j_a\})\} )]= [\bL_i, \mathbf{Y}_m] \frac{\partial \mathbf{F}(\{\mathbf{Y}_n (\{\vec{p}_a, j_a\})\} )}{\partial \mathbf{Y}_m}=0.
\end{equation}
In other words, the functions $\mathbf{Y}_n$  in  \eqref{Yn} form a basis for the Taylor expansion of any function that commute with $\bL_i$.

\paragraph{Corollary.} For $n=2$, $\mathbf{Y}_2(\vec{p})=\mathbf{X}^\dag_{\vec{p}}$. Explicitly, if $\mathbf{F}=\mathbf{F}(\mathbf{X}^\dag_{\vec{p}})$ where $\mathbf{X}_{\vec{p}}$ is defined in \eqref{N-X}, then $[\bL_i, \mathbf{F}]=0$. To prove this we note that,
\begin{equation}\begin{split}
    [\bL_i, \mathbf{F}(\mathbf{X}^\dag_{\vec{p}})]  \propto \sum_{\vec{k}}\ k_i\ [\mathbf{N}_{\vec{k}},\ \mathbf{F}(\mathbf{X}^\dag_{\vec{p}})]= p_i (\mathbf{X}^\dag_{\vec{p}}-\mathbf{X}^\dag_{-\vec{p}})\ \mathbf{F}' =0,
\end{split}
\end{equation}
where $\mathbf{F}'$ is derivative of $\mathbf{F}$ w.r.t. its argument $\mathbf{X}^\dag_{\vec{p}}$. In the last equality we used the fact that $\mathbf{X}^\dag_{\vec{k}}=\mathbf{X}^\dag_{-\vec{k}}$. 

Given the above propositions, one can construct all \textbf{Class 0} physical excited states. Here we make some facilitating remarks:

(I) The labels $m_I, w_j^I$ {of the} zero-excitation states remain unchanged under  the action of $\mathbf{F}((\bC^I_{\vec{k}})^\dag)$. Therefore,
\begin{equation}\label{m-tildem-orthogonality}
\langle \tilde{{\bf F}}; \tilde{m}_I, \tilde{w}^I_j|{\bf F}; m_I, w^I_j\rangle\propto \delta_{\tilde{m}_I, m_I}\delta_{\tilde{w}_j^I, w^I_j},   
\end{equation} independently of $\mathbf{F}, \tilde{\mathbf{F}}$. 

(II) Recalling that $\bL_i$ and $\mathbf{K}_{\vec{n}}$ besides the identity operator are only involving $\bC^I_{\vec{k}}, (\bC^I_{\vec{k}})^\dag$ operators,  the above also implies:
\begin{equation}
    \langle \tilde{{\bf F}}; \tilde{m}_I, \tilde{w}^I_j| \bL_i |{\bf F}; m_I, w^I_j\rangle\propto \delta_{\tilde{m}_I, m_I}\delta_{\tilde{w}^I_j, w^I_j}, \qquad \langle \tilde{{\bf F}}; \tilde{m}_I, \tilde{w}^I_j| \mathbf{K}_{\vec{n}} |{\bf F}; m_I, w^I_j\rangle\propto \delta_{\tilde{m}_I, m_I}\delta_{\tilde{w}^I_j, w^I_j}. 
\end{equation}

(III) For physical Hilbert spaces  consisting only of \textbf{Class 0} states, and recalling the structure of $\bL_i$ \eqref{Li-C} which consists of two terms, one {term} proportional to $m_I w_j^I$ times the identity operator and the other term proportional to number operator $\mathbf{N}_{\vec{k}}$, for a generic eigenstate of $\bL_i$ of the form \eqref{generic-excited} satisfies
\begin{equation}\label{generic-Li-eigenstate}
\bL_i |{\bf F}; m_J, w_k^J\rangle = (\sum_J m_J w^J_i+ {\cal F}_i) |{\bf F}; m_J, w^J_k\rangle,
\end{equation}
where ${\cal F}_i$ are \textit{independent} of $m_I, w^I_j$ and  depend only on the excitation numbers. In section \ref{sec:Class0ZES} we discussed the special case of ${\cal F}_i=0$. For generic \textbf{Class 0} states, however
$\sum_I m_I w^I_j=-{\cal F}_i\neq 0$. This indicates that generic excited states are not excitations of a physical ZES, {for} which  $\sum_I m_I w^I_j=0$. 

The above discussions and propositions can be summarised as follows:
\paragraph{\textbf{Class 0} Hilbert space ${\cal H}_{\text{phys}}^{(0)}$.} All physical states among \textbf{Class 0} states belong to the Hilbert space  ${\cal H}_{\text{phys}}^{(0)}$, given by:
\begin{equation}\label{H0-physical}
{\cal H}_{\text{phys}}^{(0)}=\bigcup_{m_I,w^I_j} {\cal H}^{\text{phys}}_{m_I,w^I_j}.
\end{equation}
where ${\cal H}^{\text{phys}}_{m_I,w^I_j}$ is the subspace of zero-eigenstates of $\bL_i$  for a given $m_I, w^I_j$.  Since the physicality constraints \eqref{sandwich-constraints} are bi-linear in $|\Psi\rangle$ and $|\Psi'\rangle$, it is apparent that any linear combination of  states in ${\cal H}_{\text{phys}}^{(0)}$ is also physical. Note also that as implied by \eqref{m-tildem-orthogonality}, each subspace ${\cal H}^{\text{phys}}_{m_I,w^I_j}$ component is a consistent physical Hilbert space.

\subsection{Beyond \textbf{Class 0} States }\label{sec:p2D4-Class-not0}

In this section we present {an} explicit construction of \textbf{Class 1} and \textbf{Class 2} states for null membranes in a four dimensional target space. We first discuss  \textbf{Class 2} states {followed by}  \textbf{Class 1} states. 

\subsubsection{\textbf{Class 2} States }\label{sec:p2D4-Class-2}

Consider the four mutually commuting operators 
\begin{equation}\label{Lij-2p4D}
    \bL_i^{I}:= m_I w^I_i + \sum_{\vec{p}} \ p_i \bC^{I\dag}_{\vec{p}}\bC^I_{\vec{p}},\qquad I,i=1,2.
 \end{equation}
Our general discussions in section \ref{sec:Class-p} is based on constructing standing waves \eqref{standing wave -Li} and the existence of matrices $(T^a)_A^i$ \eqref{TaiA}. For the $p=2, D=4$ case we have four $\bL_i^I$, two of which appear in $\bL_i$ while the remaining two are used to form $\ell^a$, as discussed in section  \ref{sec:Class-p} and \eqref{TaiA}  Below, we adopt a convenient choice for these combinations (though not the only possible one; see section \ref{sec:counting-revisited}):  
\begin{equation}\label{sep}
    \bL_1= \bL_1^{1}+\bL_1^{2},\qquad \bL_2= \bL_2^{1}+\bL_2^{2},\qquad \bJ_1:= \bL_1^{1}+\bL_2^{2}, \qquad \bJ_2:= \bL_1^{2}-\bL_2^{1}.
\end{equation}
The eigenstates of these operators are denoted by  $ |\ell_i; \jmath_i\rangle$, {satisfying:}
\begin{equation}\label{LiJi--eigenstate}
    \bL_i |\ell_i; \jmath_i\rangle= \ell_i |\ell_i; \jmath_i\rangle,\qquad  \bJ_i |\ell_i; \jmath_i\rangle= \jmath_i |\ell_i; \jmath_i\rangle.
\end{equation}
We first solve  the $\bL_i$ constraints and then  analyse  {the} $\bK_{\vec{n}}$ constraints.

\paragraph{Solutions to {$\bL_i$} constraints.} Following {the} analysis of section \ref{sec:Class-p}, \textbf{Class 2} solutions to the $\bL_i$ constraints correspond to  standing waves solutions of the form:
\begin{equation}\label{standing-wave-p2D4}
\begin{split}
|\psi\rangle_S &=\sum_{\ell_i,\jmath_i}\psi(|\ell_i|;\jmath_i)\ ||\ell_i|;\jmath_i\rangle_S\\
||\ell_i|;\jmath_i\rangle_S &:=\frac{1}{2}(||\ell_1|,|\ell_2|;\jmath_i\rangle+|-|\ell_1|,|\ell_2|;\jmath_i\rangle+||\ell_1|,-|\ell_2|;\jmath_i\rangle+|-|\ell_1|,-|\ell_2|;\jmath_i\rangle).
\end{split}
\end{equation}

\paragraph{Solutions to {$\bK_{\vec{n}}$} constraints.} As in section \ref{sec:Class-p}, we construct solutions to $\bK_{\vec{n}}$ constraints among  (certain linear combinations) of  $||\ell_1|,|\ell_2|;\jmath_i\rangle_S $ states.  To proceed,  we note  the commutation relations:
\begin{subequations}\label{LiJi-Kn-commutation}
\begin{align}
   [\bL_i,\bK_{\vec{n}}]&=- n_i\bK_{\vec{n}}, \label{Li-Kn-commutation}\\ 
   [\bJ_1,\bK_{\vec{n}}]&=- n_i\bK^{i}_{\vec{n}}, \label{H1-Kn-commutation}\\  
    [\bJ_2,\bK_{\vec{n}}]&=- \epsilon_{ij} n_i\bK^{j}_{\vec{n}},\label{H2-Kn-commutation}
\end{align}
\end{subequations}
where we used \eqref{LiI-Krn-algebra},  with $\bK_{\vec{n}}=\bK^{1}_{\vec{n}}+\bK^{2}_{\vec{n}}$ and $\bK^{i}_{\vec{n}}$ are defined in \eqref{K-n-C-m}.

Sandwiching the  commutations between two  eigenstates  $|\ell_i; \jmath_i\rangle$ yields: 
\begin{subequations}\label{LiJi-Kn-sandwich}
\begin{align}
 (\tilde{\ell}_i-\ell_i+ n_i) \langle \tilde{\ell}_i; \tilde{\jmath}_i |\bK_{\vec{n}}|\ell_i; \jmath_i\rangle &=0, \label{Li-Kn-sandwich}\\ 
  (\tilde{\jmath}_1-\jmath_1+ n_1) \langle \tilde{\ell}_i; \tilde{\jmath}_i |\bK^{1}_{\vec{n}}|\ell_i; \jmath_i\rangle &+ (\tilde{\jmath}_1-\jmath_1+ n_2)\langle \tilde{\ell}_i; \tilde{\jmath}_i |\bK^{2}_{\vec{n}}|\ell_i; \jmath_i\rangle=0, \label{J1-Kn-sandwich}\\  
    (\tilde{\jmath}_2-\jmath_2- n_2) \langle \tilde{\ell}_i; \tilde{\jmath}_i |\bK^{1}_{\vec{n}}|\ell_i; \jmath_i\rangle &+ (\tilde{\jmath}_2-\jmath_2+ n_1)\langle \tilde{\ell}_i; \tilde{\jmath}_i |\bK^{2}_{\vec{n}}|\ell_i; \jmath_i\rangle=0.\label{J2-Kn-sandwich}
\end{align}
\end{subequations}
If $\tilde{\ell}_i\neq \ell_i$ then \eqref{Li-Kn-sandwich} does not yield any condition on $\langle \tilde{\ell}_i; \tilde{\jmath}_i |\bK_{\vec{n}}|\ell_i; \jmath_i\rangle$. From \eqref{J1-Kn-sandwich} and \eqref{J2-Kn-sandwich} we learn that $\langle \tilde{\ell}_i; {\jmath}_i |\bK^{1}_{\vec{n}}|\ell_i; \jmath_i\rangle=\langle \tilde{\ell}_i; {\jmath}_i |\bK^{2}_{\vec{n}}|\ell_i; \jmath_i\rangle=0$, i.e. if $\jmath_i=\tilde{\jmath}_i$ then $\langle \tilde{\ell}_i; {\jmath}_i |\bK_{\vec{n}}|\ell_i; \jmath_i\rangle=0$ for any $\vec{n}$. Since standing waves \eqref{standing-wave-p2D4} are eigenstates of $\bJ_i$, then ${}_S\langle \jmath_i; |\tilde{\ell}_i|| \bK_{\vec{n}}||\ell_i|;\jmath_i\rangle_S=0$. 

Finally, we need to make sure that the operator product  requirement \eqref{Operator-Product} is also fulfilled. Consider operators ${\cal O}$ and $\tilde{\cal O}$ that are respectively associated with  $||\ell_i|;\jmath_i\rangle_S$ and $||\tilde{\ell}_i|;\tilde{\jmath}_i\rangle_S$. The state associated with ${\cal O}\tilde{\cal O}$ is then
\begin{equation}
    ||\ell_i+\tilde\ell_i|;\jmath_i+\tilde{\jmath}_i\rangle_S+||\ell_i-\tilde\ell_i|;\jmath_i+\tilde{\jmath}_i\rangle_S
\end{equation}
Thus \eqref{Operator-Product} is fulfilled iff  $\jmath_i=\tilde{\jmath}_i=0$. All in all, our physical \textbf{Class 2} states are of the form
\begin{equation}\label{Class2-p2D4}
|\textbf{2}; \psi(|\ell_i|)\rangle_S =\sum_{\ell_i}\ \psi(|\ell_i|)\ ||\ell_i|;\jmath_i=0\rangle_S. 
\end{equation}
With the above general construction and in particular \eqref{Class2-p2D4} in hand, we can now construct \textbf{Class 2} states within the total Hilbert space ${\cal H}$ constructed in section \ref{sec:tot-Hil}. We remark, however, that the choices for $\bJ_1, \bJ_2$ combinations are not limited to those in \eqref{sep}. As we will discuss in section \ref{sec:counting-revisited} there are many other choices for these two operators that yield physically inequivalent super-selection sectors within \textbf{Class 2}. 

\paragraph{\textbf{Class 2} ZES.}
For a ZES $|0; m_I, w^I_j\rangle$, 
\begin{equation}\label{ZES-ell-jmath}
    \ell_1= m_1 w^1_1+ m_2 w^2_1,\qquad \ell_2= m_1 w^1_2+ m_2 w^2_2,\qquad \jmath_1= m_1 w^1_1+m_2 w^2_2,\qquad \jmath_2= m_2 w^2_1-m_1 w^1_2,
\end{equation}
For a \textbf{Class 2} state, in $\jmath_i=0$ sector, we have $m_2 w^2_1=m_1 w^1_2, m_2 w^2_2=-m_1 w^1_1$ and 
\begin{equation}\label{ZES-ell1-ell2-Class2}
    \ell_1= m_1 (w^1_1+ w^1_2),\qquad \ell_2= m_1 (w^1_2- w^1_1),
\end{equation}
and that $\ell_i\neq 0$. One can then write all \textbf{Class 2} ZES as in \eqref{standing-wave-p2D4} and \eqref{Class2-p2D4}.

\paragraph{Excited \textbf{Class 2} states.} 

Recalling \eqref{Lij-2p4D}, which may be written as
\begin{equation}\label{Lij-2p4D-derivative}
    \bL_i^{I}:=m_I w^I_i +\sum_{\vec{p}} \ p_i \bC^{I\dag}_{\vec{p}}\frac{\partial}{\partial\bC^{I\dag}_{\vec{p}}},
 \end{equation}
for a generic monomial excited state  $|{\bf F}; m_I, w^I_j\rangle$ \eqref{generic-excited} and \eqref{level-n-monomial-excitation} 
\begin{equation}\label{Lij-2p4D-excitation}
    \bL_i^{I} |{\bf F}; m_I, w^I_i\rangle =\big(m_I w^I_i\ {\bf F}+ \sum_{\vec{p}} \ p_i \bC^{I\dag}_{\vec{p}}\frac{\partial {\bf F}}{\partial\bC^{I\dag}_{\vec{p}}}\big)|0; m_I, w^I_i\rangle:=\big(m_I w^I_i+ \sum_{\vec{p}} \ p_i N^I_{\vec{p}}\big)|{\bf F}; m_I, w^I_i\rangle,
 \end{equation}
where $N^I_{\vec{p}}$ counts number of $\bC^{I\dag}_{\vec{p}}$ in ${\bf F}$. For these states, 
\begin{equation}\label{ellij-Njp}
\begin{split}
    \ell_i^I= m_I w^I_i &+ \sum_{\vec{p}} \ p_i N^I_{\vec{p}}\\
    \ell_1=\ell_1^1+\ell_1^2,\qquad  \ell_2=\ell_2^1+\ell_2^2,&\qquad  \jmath_1= \ell_1^1+\ell_2^2,\qquad \jmath_2= \ell_1^2-\ell_2^1.
\end{split}
\end{equation}
The requirement of $\jmath_1=\jmath_2=0$ yields,
\begin{equation}\label{ell1-ell2-Class2}   
    \ell_1=\ell_1^1+\ell_1^2,\qquad  \ell_2=-\ell_1^1+\ell_1^2,\qquad \ell_2^2=-\ell_1^1,\qquad \ell_2^1=\ell_1^2,
\end{equation}
With the above we can explicitly write \eqref{standing-wave-p2D4} and \eqref{Class2-p2D4}. 

\subsubsection{\textbf{Class 1} States}\label{sec:class-1-p2D4}
For \textbf{Class 1} states, there are two possible choices: either $\ell_1=0$ xor $\ell_2=0$. Here we focus on the case $\ell_2=0$ and keep $\ell_1\neq 0$, the alternative case is analogous and will not be elaborated upon here. With this choice, the $\bL_2$ constraint is readily satisfied  as we are dealing with states of the form $|\ell_1, \ell_2=0; \jmath_1, \jmath_2\rangle$. The $\bL_1$ constraints are resolved through the standing wave solutions: 
\begin{equation}\label{Class1-standing-wave}
||\ell_1|,\ell_2=0;\jmath_1,\jmath_2\rangle_S=\frac{1}{\sqrt{2}}(||\ell_1|,\ell_2=0;\jmath_1,\jmath_2\rangle+|-|\ell_1|,\ell_2=0;\jmath_1,\jmath_2\rangle.
\end{equation}
As we commented in the \textbf{Class 2} case, the possible choices for $\bJ_1, \bJ_2$ are limited to those specified in \eqref{sep}. These possibilities will be further explored in section \ref{sec:counting-revisited}.

To solve $\bK_{\vec{n}}$ constraint for \textbf{Class 1} states, we use \eqref{LiJi-Kn-sandwich} which is written for any two generic states $|\ell_1, \ell_2; \jmath_1, \jmath_2\rangle$. For $\ell_2=\tilde\ell_2=0$,  \eqref{Li-Kn-sandwich}  with $i=2$  reduces to  
\begin{equation}
 n_2\  \langle \tilde{\ell}_1,\tilde{\ell}_2=0; \tilde{\jmath}_1,\jmath_2\ |\bK_{\vec{n}}|\ell_1,\ell_2=0; \jmath_1,\jmath_2\rangle=0, 
\end{equation}
Thus, $\bK_{\vec{n}}$ constraint is satisfied for {all} state with $\ell_2=0$ {when $n_2\neq 0$}. To explore {the case} $n_2=0$ and $n_1\neq 0$\footnote{Note that $\bK_{\vec{n}}$ is only defined for $\vec{n}\neq 0$ and one can choose $\bK_{\vec{n}=0}=0$.} consider \eqref{J1-Kn-sandwich} and \eqref{J2-Kn-sandwich}, which read as 
\begin{subequations}\label{Ji-Kn-sandwich-n2-0}
\begin{align}
 (\tilde{\jmath}_1-\jmath_1+ n_1) \langle \tilde{\ell}_i; \tilde{\jmath}_i |\bK^{1}_{\vec{n}}|\ell_i; \jmath_i\rangle + (\tilde{\jmath}_1-\jmath_1)\langle \tilde{\ell}_i; \tilde{\jmath}_i |\bK^{2}_{\vec{n}}|\ell_i; \jmath_i\rangle&=0, \label{J1-Kn-n2-0sandwich}\\  
    (\tilde{\jmath}_2-\jmath_2) \langle \tilde{\ell}_i; \tilde{\jmath}_i |\bK^{1}_{\vec{n}}|\ell_i; \jmath_i\rangle + (\tilde{\jmath}_2-\jmath_2+ n_1)\langle \tilde{\ell}_i; \tilde{\jmath}_i |\bK^{2}_{\vec{n}}|\ell_i; \jmath_i\rangle&=0.\label{J2-Kn-n2-0sandwich}
\end{align}
\end{subequations}
Adding up the above equations we find 
\begin{equation}\label{Kn-sandwich-n2-0-summed}
     (\tilde{\jmath}_1+\tilde{\jmath}_2-\jmath_1-\jmath_2+ n_1) \langle \tilde{\ell}_i; \tilde{\jmath}_i |\bK_{\vec{n}}|\ell_i; \jmath_i\rangle=0.
\end{equation}
Since $n_1\neq0$ \eqref{Kn-sandwich-n2-0-summed} holds only if  $\tilde{\jmath}_1+\tilde{\jmath}_2=\jmath_1+\jmath_2$. The requirement \eqref{Operator-Product} then implies that we should work with states with $\jmath_1+\jmath_2=0$. If we choose $\ell_1=0, \ell_2\neq 0$ case, the above analysis  yields $\jmath_1=\jmath_2$.

To summarize, all \textbf{Class 1} states have either $\ell_1=0, \ell_2\neq 0$ or $\ell_2=0, \ell_1\neq 0$ and for these choices we need to respectively take $\jmath_1=\jmath_2$ or $\jmath_2=-\jmath_1$. Among the two choices, we showcase $\ell_1=0=\jmath_1-\jmath_2$ here for which the most general \textbf{Class 1} state is of the form 
\begin{equation}\label{Class1-p2D4}
|\textbf{1}; \psi(|\ell|,\jmath)\rangle_S =\sum_{\ell,\jmath}\ \psi(|\ell|,\jmath)\ |\ell_1=0 ,|\ell|;\jmath_1=\jmath,\jmath_2=\jmath \rangle_S,
\end{equation}
with $|0, |\ell|;\jmath,\jmath\rangle_S$ defined in \eqref{Class1-standing-wave}.  As already mentioned, besides the two $\bJ_1, \bJ_2$ choices in \eqref{sep}, one could  define other linear combinations of $\bL_i^I$ as $\bJ_i$'s which yield different \textbf{Class 1} super-selection sectors. This will be discussed in section \ref{sec:counting-revisited}.

\paragraph{\textbf{Class 1} ZES.}
For ZES  $|0; m_I, w_j^I\rangle$, recalling \eqref{ZES-ell-jmath},  $\ell_2=0=\jmath_1+\jmath_2$ implies
\begin{equation}
   \ell_1= 2 m_1 w_2^1,\qquad \jmath_1= m_1 (w^1_1- w^1_2),\qquad  m_1 (w_1^1-w^2_1)= m_2 (w^2_2-w^2_1),\qquad  m_2 w^2_2=-m_1 w^1_2.
\end{equation}
 The generic state is as given in \eqref{Class1-p2D4}.

\paragraph{Excited \textbf{Class 1} states.}
The discussion  here closely parallels that of the generic \textbf{Class 2} excited state. Starting from \eqref{ellij-Njp}, we should now impose
\begin{equation}\label{ell1-jmath1-Class1}   
\begin{split}
    \ell_1:=\ell=\ell_1^1+\ell_1^2,\qquad  \jmath_1=-\jmath_2:=&\jmath=\ell_1^1+\ell_2^2=\ell_2^1-\ell_1^2,\qquad \ell_2=\ell_2^2+\ell_1^2=0. \\ 
&\text{or}\\
    \ell_1=\ell_1^1+\ell_1^2=0,\qquad  \jmath_1=\jmath_2:=\jmath=&\ell_1^1+\ell_2^2=\ell_1^2-\ell_2^1,\qquad \ell_2:=\ell=\ell_2^2+\ell_1^2,     
\end{split}
\end{equation}
With these conditions, we can explicitly write \eqref{Class1-standing-wave} and \eqref{Class1-p2D4}. We note that the two choices above are mapped to each other upon worldvolume parity under which $\ell^I_1\leftrightarrow \ell^I_2$.

\subsection{Further Comments on 4D Membrane Physical States }\label{sec:counting-revisited}

We showed that one can recognize three \textbf{Class } $\bcN=0,1,2$ super-selection sectors and that each is labelled by two integers, which are two linearly independent combinations of $\ell_i^I$: 
\begin{itemize}
    \item \textbf{Class 0} states, $\ \jmath_1=\ell_1^1-\ell^1_2,\ \jmath_2=-(\ell_1^1+\ell^1_2)$ with $\ell_2^2=-\ell^1_2, \ \ell^2_1=-\ell_1^1$.
    \item \textbf{Class 1} states, $\ |\ell_2|=|\ell_2^1+\ell^2_2|,\ \jmath_1=\jmath_2=\ell_1^1+\ell^2_2=\ell_1^2-\ell^1_2$ with $\ell_2^2=-\ell^1_2$.
    \item \textbf{Class 2} states: $\ |\ell_1|=|\ell_1^1+\ell^2_1|=|\ell_2^2-\ell_2^1|,\ |\ell_2|=|\ell_2^1+\ell^2_2|=|\ell_1^1-\ell^2_1|$ with $\ell_2^2=-\ell^1_1, \ell^2_1=\ell^1_2$.
\end{itemize}
As our explicit construction shows, there are many excited states sharing the same $\ell_i^I$ labels. We note that while we have one choice for \textbf{Class 0} and \textbf{Class 2} cases, and 2 choices for \textbf{Class 1} case that are related by worldvolume parity.  A depiction of these states with the focus on $\ell_1, \ell_2$ eigenvalues is depicted in Fig.~\ref{Fig-1}.

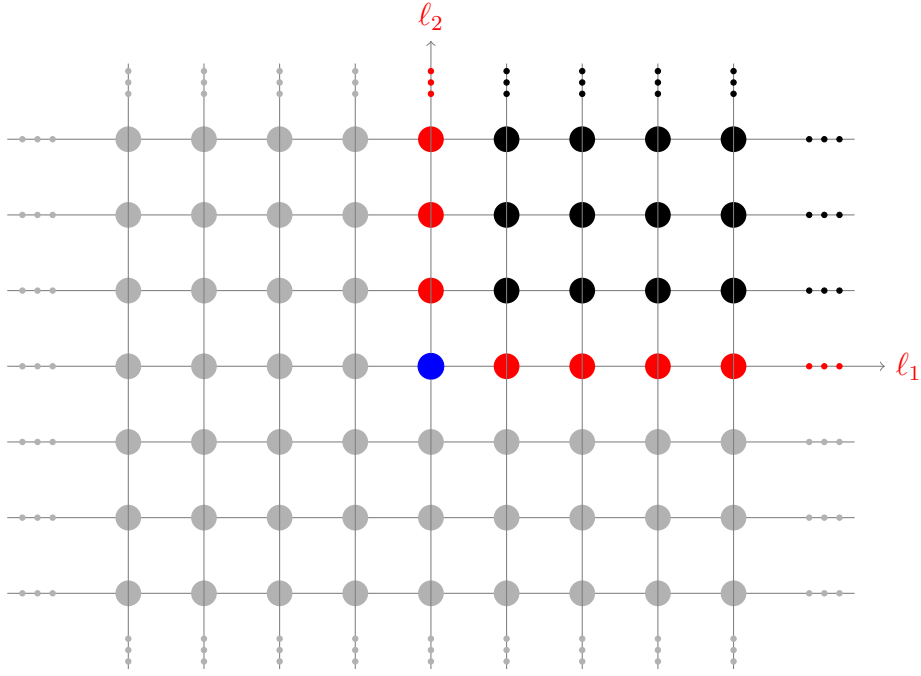
\begin{figure}[t]
\centering
\begin{tikzpicture}
\draw[gray, thin] (0.4,0) -- (11.6,0);
\filldraw[black!30] (.6,0) circle (1pt);
\filldraw[black!30] (.8,0) circle (1pt);
\filldraw[black!30] (1,0) circle (1pt);
\fill[black!30] (2, 0) circle (0.17);
\fill[black!30] (3, 0) circle (0.17);
\fill[black!30] (4, 0) circle (0.17);
\fill[black!30] (5, 0) circle (0.17);
\fill[black!30] (6, 0) circle (0.17);
\fill[black!30] (7,0) circle (0.17);
\fill[black!30] (8,0) circle (0.17);
\fill[black!30] (9,0) circle (0.17);
\fill[black!30] (10,0) circle (0.17);
\filldraw[black!30] (11,0) circle (1pt);
\filldraw[black!30] (11.2,0) circle (1pt);
\filldraw[black!30] (11.4,0) circle (1pt);

\draw[gray, thin] (0.4,-1) -- (11.6,-1);
\filldraw[black!30] (.6,-1) circle (1pt);
\filldraw[black!30] (.8,-1) circle (1pt);
\filldraw[black!30] (1,-1) circle (1pt);
\fill[black!30] (2, -1) circle (0.17);
\fill[black!30] (3, -1) circle (0.17);
\fill[black!30] (4, -1) circle (0.17);
\fill[black!30] (5, -1) circle (0.17);
\fill[black!30] (6, -1) circle (0.17);
\fill[black!30] (7,-1) circle (0.17);
\fill[black!30] (8,-1) circle (0.17);
\fill[black!30] (9,-1) circle (0.17);
\fill[black!30] (10,-1) circle (0.17);
\filldraw[black!30] (11,-1) circle (1pt);
\filldraw[black!30] (11.2,-1) circle (1pt);
\filldraw[black!30] (11.4,-1) circle (1pt);

\draw[gray, thin] (0.4,1) -- (11.6,1);
\filldraw[black!30] (.6,1) circle (1pt);
\filldraw[black!30] (.8,1) circle (1pt);
\filldraw[black!30] (1,1) circle (1pt);
\fill[black!30] (2, 1) circle (0.17);
\fill[black!30] (3, 1) circle (0.17);
\fill[black!30] (4, 1) circle (0.17);
\fill[black!30] (5, 1) circle (0.17);
\fill[black!30] (6, 1) circle (0.17);
\fill[black!30] (7,1) circle (0.17);
\fill[black!30] (8,1) circle (0.17);
\fill[black!30] (9,1) circle (0.17);
\fill[black!30] (10,1) circle (0.17);
\filldraw[black!30] (11,1) circle (1pt);
\filldraw[black!30] (11.2,1) circle (1pt);
\filldraw[black!30] (11.4,1) circle (1pt);

\draw[gray, thin] (0.4,2) -- (5.83,2);
\draw[->,gray, thin] (6.17,2) -- (12,2);
\draw[red] (12,2) node[right] (script) {$\ell_1$};
\filldraw[black!30] (.6,2) circle (1pt);
\filldraw[black!30] (.8,2) circle (1pt);
\filldraw[black!30] (1,2) circle (1pt);
\fill[black!30] (2, 2) circle (0.17);
\fill[black!30] (3, 2) circle (0.17);
\fill[black!30] (4, 2) circle (0.17);
\fill[black!30] (5, 2) circle (0.17);
\filldraw[blue] (6, 2) circle (0.17);
\fill[red] (7,2) circle (0.17);
\fill[red] (8,2) circle (0.17);
\fill[red] (9,2) circle (0.17);
\fill[red] (10,2) circle (0.17);
\filldraw[red] (11,2) circle (1pt);
\filldraw[red] (11.2,2) circle (1pt);
\filldraw[red] (11.4,2) circle (1pt);

\draw[gray, thin] (0.4,3) -- (11.6,3);
\filldraw[black!30] (.6,3) circle (1pt);
\filldraw[black!30] (.8,3) circle (1pt);
\filldraw[black!30] (1,3) circle (1pt);
\fill[black!30] (2, 3) circle (0.17);
\fill[black!30] (3, 3) circle (0.17);
\fill[black!30] (4, 3) circle (0.17);
\fill[black!30] (5, 3) circle (0.17);
\fill[red] (6, 3) circle (0.17);
\fill[black!100] (7,3) circle (0.17);
\fill[black!100] (8,3) circle (0.17);
\fill[black!100] (9,3) circle (0.17);
\fill[black!100] (10,3) circle (0.17);
\filldraw[black] (11,3) circle (1pt);
\filldraw[black] (11.2,3) circle (1pt);
\filldraw[black] (11.4,3) circle (1pt);

\draw[gray, thin] (0.4,4) -- (11.6,4);
\filldraw[black!30] (.6,4) circle (1pt);
\filldraw[black!30] (.8,4) circle (1pt);
\filldraw[black!30] (1,4) circle (1pt);
\fill[black!30] (2, 4) circle (0.17);
\fill[black!30] (3, 4) circle (0.17);
\fill[black!30] (4, 4) circle (0.17);
\fill[black!30] (5, 4) circle (0.17);
\fill[color=red] (6, 4) circle (0.17);
\fill[black!100] (7,4) circle (0.17);
\fill[black!100] (8,4) circle (0.17);
\fill[black!100] (9,4) circle (0.17);
\fill[black!100] (10,4) circle (0.17);
\filldraw[black] (11,4) circle (1pt);
\filldraw[black] (11.2,4) circle (1pt);
\filldraw[black] (11.4,4) circle (1pt);

\draw[gray, thin] (0.4,5) -- (11.6,5);
\filldraw[black!30] (.6,5) circle (1pt);
\filldraw[black!30] (.8,5) circle (1pt);
\filldraw[black!30] (1,5) circle (1pt);
\fill[black!30] (2, 5) circle (0.17);
\fill[black!30] (3, 5) circle (0.17);
\fill[black!30] (4, 5) circle (0.17);
\fill[black!30] (5, 5) circle (0.17);
\fill[color=red] (6, 5) circle (0.17);
\fill[black!100] (7,5) circle (0.17);
\fill[black!100] (8,5) circle (0.17);
\fill[black!100] (9,5) circle (0.17);
\fill[black!100] (10,5) circle (0.17);
\filldraw[black] (11,5) circle (1pt);
\filldraw[black] (11.2,5) circle (1pt);
\filldraw[black] (11.4,5) circle (1pt);

\draw[gray, thin] (5,-2) -- (5,6);
\draw[gray, thin] (4,-2) -- (4,6);
\draw[gray, thin] (3,-2) -- (3,6);
\draw[gray, thin] (2,-2) -- (2,6);
\draw[gray, thin] (7,-2) -- (7,6);
\draw[gray, thin] (8,-2) -- (8,6);
\draw[gray, thin] (9,-2) -- (9,6);
\draw[gray, thin] (10,-2) -- (10,6);
\draw[gray, thin] (6,-2) -- (6,1.83);
\draw[->,gray, thin] (6,2.17) -- (6,6.3);
\draw[red] (6,6.3) node[above] (script) {$\ell_2$};
\filldraw[black!30] (5,-1.9) circle (1pt);
\filldraw[black!30] (5,-1.75) circle (1pt);
\filldraw[black!30] (5,-1.6) circle (1pt);
\filldraw[black!30] (4,-1.9) circle (1pt);
\filldraw[black!30] (4,-1.75) circle (1pt);
\filldraw[black!30] (4,-1.6) circle (1pt);
\filldraw[black!30] (3,-1.9) circle (1pt);
\filldraw[black!30] (3,-1.75) circle (1pt);
\filldraw[black!30] (3,-1.6) circle (1pt);
\filldraw[black!30] (2,-1.9) circle (1pt);
\filldraw[black!30] (2,-1.75) circle (1pt);
\filldraw[black!30] (2,-1.6) circle (1pt);
\filldraw[black!30] (6,-1.9) circle (1pt);
\filldraw[black!30] (6,-1.75) circle (1pt);
\filldraw[black!30] (6,-1.6) circle (1pt);
\filldraw[black!30] (7,-1.9) circle (1pt);
\filldraw[black!30] (7,-1.75) circle (1pt);
\filldraw[black!30] (7,-1.6) circle (1pt);
\filldraw[black!30] (8,-1.9) circle (1pt);
\filldraw[black!30] (8,-1.75) circle (1pt);
\filldraw[black!30] (8,-1.6) circle (1pt);
\filldraw[black!30] (9,-1.9) circle (1pt);
\filldraw[black!30] (9,-1.75) circle (1pt);
\filldraw[black!30] (9,-1.6) circle (1pt);
\filldraw[black!30] (10,-1.9) circle (1pt);
\filldraw[black!30] (10,-1.75) circle (1pt);
\filldraw[black!30] (10,-1.6) circle (1pt);
\filldraw[black!30] (5, 5.9) circle (1pt);
\filldraw[black!30] (5,5.75) circle (1pt);
\filldraw[black!30] (5,5.6) circle (1pt);
\filldraw[black!30] (4,5.9) circle (1pt);
\filldraw[black!30] (4,5.75) circle (1pt);
\filldraw[black!30] (4,5.6) circle (1pt);
\filldraw[black!30] (3,5.9) circle (1pt);
\filldraw[black!30] (3,5.75) circle (1pt);
\filldraw[black!30] (3,5.6) circle (1pt);
\filldraw[black!30] (2,5.9) circle (1pt);
\filldraw[black!30] (2,5.75) circle (1pt);
\filldraw[black!30] (2,5.6) circle (1pt);
\filldraw[red] (6,5.9) circle (1pt);
\filldraw[red] (6,5.75) circle (1pt);
\filldraw[red] (6,5.6) circle (1pt);
\filldraw[black] (7,5.9) circle (1pt);
\filldraw[black] (7,5.75) circle (1pt);
\filldraw[black] (7,5.6) circle (1pt);
\filldraw[black] (8,5.9) circle (1pt);
\filldraw[black] (8,5.75) circle (1pt);
\filldraw[black] (8,5.6) circle (1pt);
\filldraw[black] (9,5.9) circle (1pt);
\filldraw[black] (9,5.75) circle (1pt);
\filldraw[black] (9,5.6) circle (1pt);
\filldraw[black] (10,5.9) circle (1pt);
\filldraw[black] (10,5.75) circle (1pt);
\filldraw[black] (10,5.6) circle (1pt);
\end{tikzpicture}
\caption{The 3 classes of states for a $4d$ membrane in the $(\ell_1, \ell_2)$ plane. The blue dot at the origin represents \textbf{Class 0} corresponding to the physical Hilbert space with $\ell_i = 0$.  \textbf{Class 1} states are the red dots which are either in $\ell_1>0, \ell_2=0$ and $\ell_2>0, \ell_1=0$ cases. \textbf{Class 2} states, represented by black dots, correspond to $\ell_1, \ell_2 > 0$. The gray dots belong to unphysical part of Hilbert space ${\cal H}$, which are eliminated by the sandwich constraints. As discussed, the $\bK_{\vec{n}}$ constraints impose no additional restrictions for \textbf{Class 0} states. However, for \textbf{Class 1} and \textbf{Class 2} states, the $\bK_{\vec{n}}$ constraints impose conditions on the $\jmath_1, \jmath_2$ quantum numbers, which are not depicted in this figure.}\label{Fig-1}
\end{figure}

As remarked in section \ref{sec:p2D4-Class-2}, {when} decomposing $\bL_i^{I}$ into $\bL_i, \bJ_i$, the former had a fixed form, while the choice of $\bJ_i$ had some freedom. In \eqref{sep} we chose a convenient and simple linearly independent combination.  Let us revisit this issue. A more generic choice for $\bJ_1,\bJ_2$ can be 
\begin{equation}\label{redundant}
\begin{split}
    \bL_1= \bL^{1}_1+ \bL^{2}_1 &, \qquad \bL_2=\bL^{1}_2+\bL^{2}_2\\
       \bJ_1=\sum_{I,i}\ a_I^i \bL^I_i&, \qquad \bJ_2=\sum_{I,i}\ b_I^i \bL^I_i
\end{split}
\end{equation}
where $a^i_I,b^i_I$ are 8 integer coefficients. The choice in \eqref{sep} corresponds to $a^1_1=a_2^2=1, a^1_2=a_1^2=0; b^1_1=b_2^2=0, b^1_2=-b^2_1=1$. As established above  \textbf{Class 1} (or \textbf{Class 2}) states should be zero eigenstate of one (or two) linear combination  of $\bJ_1, \bJ_2$. Vanishing of $\bJ_i$ eigenvalues depends on the choice of $a_I^i, b_I^i$. Thus, different choices for $a_I^i, b_I^i$ can lead to physically inequivalent super-selection sectors. In what follows we explore which choices for these 8 integer-valued coefficients lead to physically independent super-selection sectors. We explore \textbf{Class 2} and \textbf{Class 1} cases separately.

\subsubsection{\textbf{Class 2} case}\label{sec:class-2-counting-revisit}

To uncover the redundancies and identify the physically independent  choices of $a^i_I,b^i_I$, we recall that $\bL_i^I$ (and hence $\ell_i^I$) are defined up to $SL(2,\mathbb{Z})$ symmetry on brane coordinates, under which 
\begin{equation}
    \tilde{p}_i=\sum_j c_i{}^j\ p_j,\qquad  c_i{}^j\in \mathbb{Z},\quad \det(c)=1.
\end{equation}
Under this transformation the $\bL^I_i$'s transform as
\begin{equation}
    \tilde{\bL}^I_i=\sum_{j}c_i{}^j\bL^I_j,
\end{equation}
and therefore, 
\begin{equation}
    \tilde{\bJ}_1=\sum_{I,i}\tilde{a}_I^i \bL^I_i, \qquad \tilde{\bJ}_2=\sum_{I,i}\tilde{b}_I^i\bL^I_i,
\end{equation}
where
\begin{equation}
    \tilde{a}_I^i=\sum_j  c_j{}^i a_I^j,\qquad \tilde{b}_I^i=\sum_j c_j{}^i b_I^j .
\end{equation}
So, $\tilde{\bJ}_i$ and ${\bJ}_i$  related by $SL(2,\mathbb{Z})$ are physically equivalent. Consider the following combinations, 
\begin{equation}\label{ABC-ab}
 A \epsilon_{IJ}=\sum_{i,j}\epsilon_{ij}a_I^i a_J^j,\qquad  B \epsilon_{IJ}=\sum_{i,j}\epsilon_{ij}b_I^i b_J^j, \qquad  C_{IJ}=\sum_{i,j}\epsilon_{ij}a_I^i b_J^j,\qquad (\text{no sum on } I,J)    
\end{equation} 
where $\epsilon_{IJ}$ is an antisymmetric tensor and $\epsilon_{12}=1$. One can show that these quantities are invariant under  the transformations $a_I^i, b_I^i\to \tilde{a}_I^i, \tilde{b}_I^i$.
Thus, different choices of $a_I^i,b_I^i$ which yield different $A$, $B$ or $C_{IJ}$ cannot be transformed into one another under $SL(2,\mathbb{Z})$ and are physically distinguishable. Note that these 6 numbers are not independent, and one may show that $AB=\det(C)$, {implying} there are  5 distinct $SL(2,\mathbb{Z})$-invariant labels. One may decompose $C_{IJ}$ into symmetric and antisymmetric parts,
\begin{equation}
    C_{IJ}=C_{\{IJ\}}+ C\ \epsilon_{IJ},\qquad C_{\{IJ\}}=\frac12 (C_{IJ}+ C_{JI}),\qquad AB-C^2=\det{(C_{\{IJ\}})}. 
\end{equation}
That is, out of 8 integers  $a_I^i, b_I^i$, 3 may be removed by $SL(2,\mathbb{Z})$ and the remaining 5, {can} be labelled by $A,B$ and $C_{\{IJ\}}$.

However, not all these 5 labels are physically distinct. Recall that for  \textbf{Class 2} states $\jmath_1=\jmath_2=0$. Therefore, any two choices of $\bJ_1,\bJ_2$ which are linearly related to each other, as in 
\begin{equation}\label{lincomb}
    \bJ'_1=r_1 \bJ_1+r_2 \bJ_2, \qquad \bJ'_2=s_1 \bJ_1+s_2 \bJ_2,
\end{equation}
where $r_1 s_2-r_2 s_1\neq 0$, are physically equivalent to $\bJ_1,\bJ_2$. Thus, $a_I^i, b_I^i$ related by $GL(2,\mathbb{Z})$ transformations
\begin{equation}\label{hat-a-hat-b}
    \hat{a}_I^i=r_1 a_I^i+r_2 b_I^i,\qquad \hat{b}_I^i=s_1 a_I^i+s_2 b_I^i.
\end{equation}
are physically equivalent. To {identify} which of the 5 $SL(2,\mathbb{Z})$-invariant labels are identified by \eqref{hat-a-hat-b},  one may construct $\hat{A}, \hat{B}, \hat{C}_{IJ}$ in terms of $ \hat{a}_I^i,  \hat{b}_I^i$ in the same way ${A}, {B}, {C}_{IJ}$ {are} constructed in \eqref{ABC-ab}. One  then readily observes that,
\begin{equation}\label{hatABC}
    \begin{split}
        \hat{A}&=r_1^2 A+ r_2^2 B+2r_1r_2 C \\
         \hat{B}&=s_1^2 A+ s_2^2 B+2s_1s_2 C \\
          \hat{C}&=r_1 s_1 A+ r_2 s_2 B+(r_1s_2+r_2s_1) C \\
           \hat{C}_{\{IJ\}}&=(r_1s_2-r_2s_1) C_{\{IJ\}} 
    \end{split}
\end{equation}
As in the previous case,  $\hat{A}\hat{B}=\det{\hat{C}}=\det{(\hat{C}_{\{IJ\}})}+ \hat{C}^2$. We should then identify all $A,B,C_{IJ}$ and $\hat{A}, \hat{B}, \hat{C}_{IJ}$ that are related as in \eqref{hatABC}. One can use the first three lines in \eqref{hatABC} to set $A=B=0$ and $C=1$ and the last line may be {used} to set $\det{(C_{\{IJ\}})}=-1$. In summary, out of 5 $SL(2,\mathbb{Z})$-invariant integer-valued labels, {two} remain which can be conveniently  parametrised as
\begin{equation}\label{AB0C}
    A=B=0,\qquad C=\begin{pmatrix}
        \alpha & \beta+1 \\ \beta-1 & \gamma
    \end{pmatrix},\qquad \beta^2=\alpha\gamma+1, \qquad \alpha,\beta,\gamma\in \mathbb{Z}
\end{equation}
In addition, $\bJ_i, \bL_i$, should be linearly independent, implying that,
\begin{equation}
     \det\begin{pmatrix}
    1 & 1 & 0 & 0 \\ 0 & 0 & 1 & 1 \\ a^{1}_1 & a_{2}^1 & a^{2}_1 & a^{2}_2 \\ b^{1}_1 & b^{1}_2 & b^{2}_1 & b^{2}_2
    \end{pmatrix}=C_{12}+C_{21}-C_{11}-C_{22} \neq 0 \ \Longrightarrow 2\beta\neq \alpha+\gamma \,.
\end{equation}

We are now ready to revisit  the $\bK_{\vec{n}}$ constraint and the discussions in section \ref{sec:p2D4-Class-2}. With the $\bJ_i$ in \eqref{redundant}, we have
\begin{equation}
    [\bJ_1,\bK^{I}_{\vec{n}}]=-\sum_i n_i a_{I}^i \bK^{I}_{\vec{n}}, \qquad [\bJ_2,\bK^{I}_{\vec{n}}]=-\sum_i n_i b_{I}^i\bK^{I}_{\vec{n}}.
\end{equation}
Therefore following our previous discussions fixing $\bJ_1=\bJ_2=0$ solves the $\bK_{\vec{n}}$ constraint if for each $\vec{n}$ and $I$ either or both of $\sum_i a_{I}^i n_i$, $\sum_i b_{I}^i n_i$ are non-zero. This is guaranteed if these are linearly independent and therefore the following determinants are non-zero:
\begin{equation}
     \det \begin{pmatrix}
        a_{I}^1 & a_{I}^2 \\ b_{I}^1 & b_{I}^2
    \end{pmatrix}= \sum_{i,j} \epsilon_{ij}\ a_{I}^i b_{I}^j =C_{II} \neq 0 \ \ \Longrightarrow \alpha,\gamma\neq 0.
\end{equation}
To summarize, besides $|\ell_i|$ there are two other integer labels $\alpha,\gamma$, where $\alpha, \gamma\neq 0, \alpha\neq \gamma\pm 2, 1+\alpha\gamma=\beta^2, \beta\in \mathbb{Z}$, to specify the \textbf{Class 2} super-selection sectors. 

\subsubsection{\textbf{Class 1} Case}\label{sec:class-1-counting-revisit}

For this case, one may set $\ell_1=0$, while $\ell_2\neq 0$. We identified two such \textbf{Class 1} states, with $\jmath_1=\jmath_2:=\jmath$. As in the \textbf{Class 2} case,  there may be other physically independent choices for $\jmath_i$, which we explore exploiting the ideas discussed in section \ref{sec:class-2-counting-revisit}. Let $\bJ$ which we fix to zero in \textbf{Class 1} be another combination of $\bL^I_i$, different from   the one in \eqref{sep}:
\begin{equation}
    \bJ=\sum a^i_I \bL^I_i,\qquad \bL_1= \sum b^i_I \bL^I_i:=\bL^{1}_1+\bL^{2}_1.
\end{equation}
That is, $\bL_1$, $b^2_1=b^2_2=0, b^1_1=b^1_2=1$ and for the solution discussed in section \ref{sec:class-1-p2D4} $a_1^1=a_2^2=a_2^1=-a_1^2=1$. There are initially 4 integers to be chosen to fix $\bJ$, but as in the case of \textbf{Class 2}, there are redundancies in this choice. As before one of them is the $SL(2,\mathbb{Z})$ symmetry on the brane coordinates. We note that by choosing $\ell_1=0$, the only remaining  part of this redundancy is scaling $ {\bL}_1$ by an integer, i.e. $b^2_1=b^2_2=0, b^1_1=b^1_2=b$.

The other redundancy stems from the freedom  in choosing $\bJ$; $\hat{\bJ}$ and $\bJ$ that are related as
\begin{equation}\label{superposition-Class-1}
    \hat{\bJ}=r \bJ+ s \bL_1, 
\end{equation}
are physically the same.   To fix this redundancy, as in \eqref{ABC-ab} one may calculate $A, B, C_{IJ}$
\begin{equation}
   A=\frac12 \sum_{i,j}\ \epsilon^{IJ}\epsilon_{ij}a^i_I a^j_J,\qquad B=\frac12 \sum_{i,j}\ \epsilon^{IJ}\epsilon_{ij}b^i_I b^j_J=0, \qquad C_{IJ}=\sum_{i,j}\epsilon_{ij}a^i_I b^j_J=a^1_I b^2_J,
\end{equation}
and hence $C_{12}=C_{11}=a^1_1b, C_{21}=C_{22}=a^1_2b$. Eq.~\eqref{superposition-Class-1} then implies that hatted and un-hatted quantities,  
\begin{equation}
    \hat A=r(r A+s C),\qquad \hat C= r C,\qquad \hat{C}_{\{IJ\}}=r C_{\{IJ\}},
\end{equation}
are physically equivalent and should be identified. These imply $a_2^1=a_1^1, a_2^2=-a_1^2$. The $\bK_{\vec{n}}$ constraints require $a_1^1, a_2^2\neq 0$. Thus, super-selection sectors are  labelled by two non-zero integers which may be conveniently chosen to be $a_1^1$ and $a_2^2$.  Explicitly,  for $\bL_2=0$ \textbf{Class 1} sector, a generic $\bJ$ which may be set to zero may be written as $\bJ=a_2^2(\bJ_1+\bJ_2)+(a_1^1-a_2^2)\bL_1$.

\section{Discussion and Outlook}\label{sec:discussion}

In this work, we tackled the  quantisation of null $p$-branes on a $D$ dimensional flat Minkowski {spacetime} ($D\geq p+2$). The worldvolume of null $p$-brane is conveniently parametrised by congruence of null geodesics passing {through} a $p$-dimensional spacelike hypersurface embedded {in} the $D-1$ dimensional constant-time slice of the target space. This embedding is described by $ X^\alpha = X^\alpha(x^\mu(\sigma^i), p^\mu(\sigma^i); \tau) $, where $ X^\mu $ are the embedding coordinates of the worldvolume, $ \sigma^i $ parameterise the $ p $-dimensional spatial sector of the worldvolume, and $ x^\mu $, $ p^\mu $ represent the initial position and momentum of each null ray within the congruence. {The functions $ x^\mu(\sigma^i) $ and $ p^\mu(\sigma^i) $ are not arbitrary but are constrained by a set of $ p+1 $ constraints, $ {\cal C}_0 = 0 $ and $ {\cal C}_i = 0 $. These constraints enforce the structure of the null congruence, transforming it into a null $ p $-brane. } Remarkably the  constraints form a BMS$_{p+2}$ algebra, {reflecting the underlying symmetry of the system}. 

We solved a {significant portion} of the BMS$_{p+2}$ constraints through fixing the light-cone gauge at classical level. In this gauge, two of {the embedding coordinates,}  $X^\pm$, are {expressed} in terms of the remaining $D-2$ {coordinates} $X^I$'. These $ X^I $ are then subject to the residual constraints that remain after gauge fixing. 
The {residual } constraints at classical level {include}  $p$ ``level matching'' conditions,  along with  $p-1$ {additional} constraints of the form $K^r(\sigma^i)=0$. {To address these}, we applied the sandwich quantisation {scheme,  applying it systematically} to {the}  remaining constraints. The problem of quantisation of the null brane, hence boils down to {problem of} consistently imposing {these}  constraints at quantum level and solving them,  ensuring the construction of a well-defined quantum theory.

\paragraph{Right action vs. sandwich quantisation schemes.} Following \cite{Bagchi:2020fpr, Bagchi:2021rfw, Bagchi:2021ban, Bagchi:2024tyq}, we argued that {the} consistency of quantisation  requires defining the physical Hilbert space such that the constraints vanish when sandwiched between any two physical states. {This approach, termed}  ``sandwich quantisation scheme''  (detailed in appendix \ref{appen:Sandwich-quantisation}) {extends}  Dirac's procedure \cite{Dirac:1996} and {is applicable }  to  systems with second class constraints. The sandwich quantisation {method} {can}  be {contrasted}  with the more conventional treatment of constraints, the ``right action'': Requiring the constraint (which is a quantum operator {acting}  on a total Hilbert space) on any  physical state be zero; or, physical states are zero eigenstates of the constraints. It is apparent that all solutions to the right action quantisation are also solutions to the sandwich conditions, but not vice-versa. The first example of such comparison has been carried out for the usual tensile string \cite{Bagchi:2024tyq}: Sandwich quantisation scheme for the tensile string case reveals ``quantum worldsheet equivalence principle'' which is not manifest in textbook treatment of string theory.

In our analysis, \textbf{Class 0} states are characterized as zero eigenstates of only a subset of the constraints, specifically the $\bL_i$, and are not zero eigenstates of the $\bK^r_{\vec{n}}$.  Indeed, none of the \textbf{Class} $\bcN$ physical states are zero eigenstates of the full set of constraints. That is, the right-action quantisation scheme for null $p$-branes in the light-cone gauge does not yield a physically  satisfactory solution; the right-action quantisation imposes constraints that are overly restrictive and needlessly tight and rigid. From our perspective, this rigidity represents a critical missing piece in the broader framework for the quantisation of branes. The sandwich quantisation scheme, on the other hand, seems to be a more promising alternative. As demonstrated explicitly in this work, it offers a more tractable and physically meaningful approach to the quantisation of $p$-branes. {While our investigation focused on the technically simpler case of null $p$-branes, the principles, methods, and techniques underpinning the sandwich quantisation scheme appear to be equally applicable to the case of tensile $p$-branes.}

\paragraph{Sandwich quantisation scheme and (partial) gauge fixing.} Building upon preceding discussions on right action vs sandwich schemes, one is led to another related important question: Does treating all the original constraints through sandwich conditions  lead to  exactly the same physics (if not exactly the same Hilbert space, see \cite{Bagchi:2024tyq} for an explicit example) as one obtains after  light-cone gauge-fixing, i.e. solving for a part of the constraints and imposing the rest through sandwich conditions?  In the case of tensile strings, this question has been thoroughly addressed in string theory textbooks: it is shown that the imposing Virasoro constraints as ``right-action,'' fixing the light-cone gauge and dealing only with level-matching condition is physically equivalent to the covariant quantisation \cite{Green:1987sp, Polchinski:1998rq}. We will explore this question in our upcoming work.

As a related remark, let us consider a system of constraints $C_i=0$ that form an algebra ${\cal G}$. One may choose to explicitly solve a subset of these constraints, say $ C_a = 0 $, and treat the remaining constraints, $ C_\alpha = 0 $, via either the sandwich or right-action schemes.  This raises two important questions: (i) Under what conditions is this procedure mathematically consistent, and (ii) does the choice of which subset of constraints to solve versus impose at the quantum level lead to differing physical results?   With respect to the first question, a necessary (though not necessarily sufficient) condition for consistency is that the remaining constraints $ C_\alpha = 0 $ form a closed algebra. In the specific case at hand, light-cone gauge fixing leads to the constraints $\bL_i$ and $\bK^r_{\vec{n}}$, which collectively define the algebra of constraints \eqref{constraint-algebra}. However, given that $\mathbf{M}^2$ commutes with these constraints, one is left with two choices for definition of the mass, as discussed in section \ref{sec:mass}.  Exploring these issues, both within the current context and in broader settings involving constrained systems, offers valuable insights into the interplay of gauge fixing, quantisation schemes, and the resulting physical interpretations. This remains a promising direction for future investigation.

\paragraph{Choice of vacuum states.} In addressing the sandwich constraints, we began with a total Hilbert space constructed by applying all possible combinations of creation operators to a vacuum state. This vacuum state is defined as the one annihilated by all annihilation operators, cf. \eqref{ZES-def}.  {An important question arises: are there different consistent choices for the vacuum state, and does the physical Hilbert space obtained by imposing the sandwich constraints critically depend on the choice of vacuum state and/or creation operators?} For the case of null string this question was discussed in some recent papers \cite{Bagchi:2020fpr, Bagchi:2020ats}. {If physically consistent alternative vacuum states are indeed possible, for example as proposed in \cite{Bagchi:2020fpr}, it becomes imperative to investigate whether these vacua are related by Bogoliubov transformations.} These questions should be explored for the case of null $p$-branes too.   

\paragraph{Physics of \textbf{Class} $\bcN$ states.} We have established that  solutions to {the} sandwich conditions \eqref{sandwich-constraints} for a null $p$-brane {are} classified {in} $p+1$ {distinct} \textbf{Class} $\bcN$ super-selection sectors. In addition, as discussed in section \ref{sec:counting-revisited}, additional super-selection sectors can be identified within each of the initial $p+1$  sectors. As already pointed out, none of these super-selection sectors appear in {the} right action quantisation scheme. This raises the intriguing possibility that these sectors correspond to distinct physical states accessible to different sets of world-volume observers, analogous to the case of strings as explored in \cite{Bagchi:2024tyq}. While the detailed physical implications and dynamics of these super-selection sectors have not been addressed in this work, their analysis is both imperative and promising for a deeper understanding of null $p$-brane physics. This remains an important direction for future investigations.

\paragraph{Relevance to black holes and the membrane paradigm.} As outlined briefly in the introduction, null branes may have a profound connection to the physics of black holes via the membrane paradigm \cite{Thorne:1986iy, Price:1986yy, Grumiller:2018scv, Grumiller:2022qhx}.  The membrane residing at the event horizon proposed in the membrane paradigm is a prime example of a null brane of the type discussed in this work. In particular, for a typical $D$ dimensional  black hole horizon is a codimension-1 null surface and hence one deals with codimension-2 branes, $p=D-2$. For the simplest case of $D=3$, one deals with null strings. This idea was  analyse d in \cite{Bagchi:2022iqb} where an explicit construction for 3 dimensional black hole microstates was given. The methods and insights developed in this paper can be naturally extended to more interesting higher dimensional black holes. In particular, the case of $D=4$,  corresponding to Kerr or Schwarzschild black holes is related to  $p=2, D=4$,  studied in detail in section \ref{sec:p2D4}.  Extending the framework and proposals of \cite{Bagchi:2022iqb} to this setting is a compelling avenue for future research, with the potential to deepen our understanding of black hole microstates and their underlying null brane structures.

\paragraph{Consistency of null $p$-brane theory.} Last but not least, here we explored free null $p$-brane theory. {Analogous to conventional string theory, it is natural to expect that the quantum consistency of $p$-brane theory may impose constraints on the dimensionality of the target spacetime and may necessitate the inclusion of supersymmetry.}   Moreover, interactions may also lead to other consistency requirements. By taking the first steps of formulating the null brane theory, we hope our work paves the way to all these physically interesting and important questions.

\section*{Acknowledgement}
We particularly thank Daniel Grumiller for many discussions and for collaboration at early stages of this work. We would like to thank Arjun Bagchi, Aritra Banerjee, Sergio Cecotti, Paul Townsend and especially Jorge Russo for  helpful discussions and comments. SD acknowledges support from  FRS-FNRS (Belgium) through the PDR/OL C62/5 project ``Black hole horizons: away from conformality'' (2022-2025) and thanks the hospitality of TU and ESI, Vienna at early stages of this project. MMShJ would like to thank ICCUB, Barcelona for the hospitality where this work was completed. The work of MMShJ is in part supported by the INSF grant No 4026712.  The work of HY is supported in part by Beijing Natural Science Foundation under Grant No. IS23013. 

\appendix

\section{Sandwich quantisation Scheme}\label{appen:Sandwich-quantisation}

quantisation starts with defining the Hilbert space that contains all physical states; {this} Hilbert space serves as the quantum analogue of the classical phase space.  Physical observables are represented as operators acting on this Hilbert space. In a quantum field theory, {when} a well-defined vacuum state exists,  the operator-state correspondence  {asserts that} a physical state {can be uniquely associated with each}  physical observable operator. Moreover, the product of two physical operators defined at the same point should be a physical operator, modulo normal ordering considerations. 

For constrained systems, the quantisation procedure should be handled with care. Dirac's seminal work \cite{Dirac1950, Dirac:1996} provides the foundational framework for addressing such cases.  In Dirac's formalism, gauge symmetries are treated as first-class constraints, while constraints with non-vanishing Poisson brackets are categorized as second-class constraints. Dirac's prescription for second-class constraints involves replacing the Poisson bracket with the Dirac bracket, which is subsequently promoted to commutators during quantisation. { First-class constraints, on the other hand, necessitate a gauge-fixing procedure.} See \cite{Henneaux:1994lbw} for details. {Constraints are generally expressed as functions on the phase space, representing relations among degrees of freedom (dof) that vanish on-shell, denoted   $ C_A \approx 0   $. The consistency of the system of constraints requires that their Poisson brackets form a closed algebra. Additionally, for any physical observable   $\mathcal{O}$, which classically corresponds to any function on the phase space, the following relations should hold:  }
\begin{equation}\label{Const-Cons-classical-PB}
    \{C_A, C_B\}_{\text{\tiny{P.B.}}} \approx 0,\qquad \{C_A, {\cal O}\}_{\text{\tiny{P.B.}}} \approx 0
\end{equation}
where $\approx$ denotes equality on-shell and on the constraint surface $C_A \approx 0$. {The Dirac bracket is constructed to ensure these relations are satisfied explicitly:  }
\begin{equation}\label{Const-Cons-classical-DB}
    \{C_A, C_B\}_{\text{\tiny{D.B.}}} = 0,\qquad \{C_A, {\cal O}\}_{\text{\tiny{D.B.}}} = 0.
\end{equation}

quantisation entails promoting functions on the phase space to operators acting on the Hilbert space. Let us denote operators associated with constraints and physical observables by $\bC_A, \bcO$ respectively. One can then envisage two possible ways of imposing the constraints at quantum level and construct physical Hilbert space.

\paragraph{Right-action quantisation scheme.} {The constraints may be imposed through the ``right action'', where the consistency conditions are expressed as: }
\begin{equation}\label{Const-Cons-quantum}
    \bC_A |\Psi\rangle =0,\qquad [\bC_A, \bC_B] |\Psi\rangle =0 ,\qquad [\bC_A, \bcO]|\Psi\rangle =0 \qquad \forall |\Psi\rangle \in {\cal H}_{\text{phys}}
\end{equation}
with ${\cal H}_{\text{phys}}$ denoting physical Hilbert space. In this framework, the operator-state correspondence implies that any physical state \(|\mathcal{O}\rangle\) can be constructed by the action of a physical observable \(\bcO\) on a vacuum state \(|\text{vac}\rangle\), where the vacuum is assumed to satisfy the physicality conditions:
\begin{equation}\label{Operator-State-Correspondence}
    \forall |{\cal O}\rangle \in {\cal H}_{\text{phys}}\ \  \ \exists \ \text{physical operator } \bcO \ \ s.t.  \ |{\cal O}\rangle=\bcO |\text{vac}\rangle. 
\end{equation}

\paragraph{Sandwich quantisation scheme.}{ The right-action conditions are often stricter than necessary. They can be relaxed and replaced with  \textit{sandwich conditions} as: }
\begin{equation}\label{Const-Cons-quantum-sandwich}
    \langle \Phi|\bC_A |\Psi\rangle =0,\qquad \langle \Phi|[\bC_A, \bC_B] |\Psi\rangle =0 ,\qquad \langle \Phi|[\bC_A, \bcO]|\Psi\rangle =0 \qquad \forall |\Phi\rangle ,|\Psi\rangle \in {\cal H}_{\text{phys}},
\end{equation}
where, for brevity, we introduce the short hand notation,  $\langle \Phi|\bC_A |\Psi\rangle \bapp \langle \bC_A \rangle$. Constraints usually form a Lie algebra, $[\bC_A, \bC_B]=f_{AB}^D \bC_D$ and hence $\langle [\bC_A, \bC_B] \rangle \bapp 0$ is not  independent from $\langle\bC_A \rangle \bapp 0$. One can show that $\langle [\bC_A, \bcO] \rangle \bapp 0$ yields $\bcO|\Psi\rangle\in {\cal H}_{\text{phys}}$. Thus, \eqref{Const-Cons-quantum-sandwich} reads as
\begin{equation}\label{Const-Cons-quantum-sandwich-II}
\tcbset{fonttitle=\scriptsize}
            \tcboxmath[colback=white,colframe=gray]{    \langle\bC_A\rangle \bapp 0,\qquad \bcO |\Psi\rangle \in  {\cal H}_{\text{phys}}  \qquad \forall |\Psi\rangle  \in {\cal H}_{\text{phys}},\quad \forall\ \text{physical observable } \bcO}
\end{equation}
From operator-state correspondence \eqref{Operator-State-Correspondence} and \eqref{Const-Cons-quantum-sandwich-II} we learn that
\begin{equation}\label{Operator-Product}
   \tcbset{fonttitle=\scriptsize}
            \tcboxmath[colback=white,colframe=gray]{ \forall\ \text{physical operators } \bcO_1, \bcO_2, \ \ \bcO_1\bc O_2 \ \text{is also a physical operator}} 
\end{equation}

\section{On Completeness of Class \texorpdfstring{$\bcN$}{} States Analysis }\label{sec:Categories-Class-II}

Suppose that our states are {labelled} by two integers $|p,q\rangle $ which are eigenstates of two commuting operators $\mathbf{P}, \mathbf{Q}$,
\begin{equation}
    \mathbf{P} |p,q\rangle= p |p,q\rangle,\qquad \mathbf{Q} |p,q\rangle= q |p,q\rangle,
\end{equation}
where $p\in \mathbb{R}$. Solutions to {the} sandwich condition $\langle \Psi|\mathbf{P} |\Phi\rangle=0$ may be constructed through linear combination of states with $p$ and $-p$, the question we want to explore is what are the options for the states with respect to $\mathbf{Q}$ operator. So let us consider a state of the form,
\begin{equation}
     |p; \psi,\phi \rangle := \sum_q \left(\psi_q |p,q\rangle + \phi_q |- p,q\rangle\right),\qquad p\geq 0
\end{equation}
Then 
\begin{equation}
     \mathbf{P} |p; \psi,\phi \rangle= p\ \sum_q \left(\psi_q |p,q\rangle - \phi_q |- p,q\rangle\right),
\end{equation}
yielding 
\begin{equation}
    \langle \tilde\phi,\tilde\psi; \tilde p| \mathbf{P} |p; \psi,\phi \rangle=  \sum_q p \delta_{p,\tilde p}\ \left( \tilde\psi_q^*\psi_q- \tilde\phi_q^*\phi_q \right) \equiv 0,\qquad \forall \tilde\phi_q, \phi_q, \tilde\psi_q, \psi_q
\end{equation}
One choice for the above to vanish is $|\psi_q|=|\phi_q|$ for any $q$, the other is generic choice is that the $q$ part of the state is fixed by its $p$ part. That is, there two categories of solutions:
\begin{itemize}
    \item \textit{Category I}: states of the form $\  |p; q \rangle+  |-p; q \rangle$,  labelled by two integers $q, p\geq 0$. 
     \item \textit{Category II}: states of the form $\  |p; q_+(p) \rangle+  |-p; {q}_-(p) \rangle$, labelled by a positive integer $p$ and two arbitrary functions of $q_\pm(p)$ such that for any two $p,\tilde{p}$, if $p=\tilde{p}$ then $q_\pm(p)=q_\pm(\tilde{p})$.
\end{itemize}
It is important to note that \textit{Category II} states cannot be written as linear combinations of a more generic looking \textit{Category I} states. 

The above {analysis can} be generalised to cases with more labels, states of the form $|p_i; q^A \rangle$ as eigenstates of operators $\mathbf{P}_i, \mathbf{Q}^A$, where $p_i, q^A$ are generic positive or negative integers. The solutions to sandwich constraints  $\langle \Psi|\mathbf{P}_i |\Phi\rangle=0$ come it different categories, states of \textit{Category I}, a basis for which are of the form
\begin{equation}
    \sum_{s_i=+,-}\ ||p|_i s_i; q^A \rangle
\end{equation}
which are labelled by non-negative integers $|p_i|$ and generic integers $q^A$, and various \textit{Category II} states of the form 
\begin{equation}
    \sum_{s_i=+,-}\ ||p|_i s_i; q_{s_i}^a(|p|_i); q^\alpha \rangle,\qquad s.t.\qquad p_i=\tilde p_i\Longrightarrow\ \tilde{q}_{s_i}^a(\tilde{|p|}_i)=q_{s_i}^a(|p|_i).
\end{equation}
where $q^A$ indices may be divided into two sets of $q^a$ and $q^\alpha$. This division is quite arbitrary and yields different \textit{Category II} sectors. In particular, $q^a$'s may be linear combinations of $|p_i|$, which are linearly independent; for this special case $\#a\leq \#i$.

However, as discussed in appendix \ref{appen:Sandwich-quantisation}, besides the sandwich conditions we need to make sure that \eqref{Operator-Product} is also satisfied. To study that, let us consider the two categories above separately. \textit{Category I} states may be written as
\begin{equation}
   |p; q \rangle+  |-p; q \rangle =({\cal O}_{p,q}+{\cal O}_{-p,q}) |\text{vac} \rangle := {\cal O}^+_{|p|,q} |\text{vac} \rangle
\end{equation}
where as the notation suggests, ${\cal O}_{p,q}$ creates a state with quantum numbers $p,q$. One may then readily verify that
\begin{equation}
    {\cal O}^+_{|p_1|,q_1} \cdot {\cal O}^+_{|p_2|,q_2}={\cal O}^+_{|p_1+p_2|,(q_1+q_2)} + {\cal O}^+_{|p_1-p_2|,(q_1+q_2)}
\end{equation}
i.e. product of two \textit{Category I} operators is a sum of two \textit{Category I} operators. So, \textit{Category I} states satisfy \eqref{Operator-Product} requirement. 

Let's next ask the same question about \textit{Category II} states, 
\begin{equation}
   |p; q_+(p) \rangle+  |-p; q_-(p) \rangle =({\cal O}_{p,q_+(p)}+{\cal O}_{-p,q_-(p)}) |\text{vac} \rangle := {\cal O}^+_{|p|;q_\pm(p)} |\text{vac} \rangle
\end{equation}
One may then readily verify that  ${\cal O}^+_{|p|,q_\pm(p)} \cdot {\cal O}^+_{|\tilde{p}|,q_\pm(\tilde{p})}$ can be written in terms of \textit{Category II} operators only if $q_\pm$ are linear functions of $p$, which basically reduces the state into a \textit{Category I} class. So, generically \textit{Category II} states do not fulfil the \eqref{Operator-Product} requirement. Therefore, in our analysis we may only focus on \textit{Category I}, as we did in the main text.



\begin{thebibliography}{10}

\bibitem{Duff:1994an}
M.~J. Duff, R.~R. Khuri, and J.~X. Lu, ``{String solitons},'' {\em Phys. Rept.}
  {\bf 259} (1995) 213--326,
  \href{http://www.arXiv.org/abs/hep-th/9412184}{{\tt hep-th/9412184}}.

\bibitem{Duff:1987cs}
M.~J. Duff, T.~Inami, C.~N. Pope, E.~Sezgin, and K.~S. Stelle, ``{Semiclassical
  quantisation of the Supermembrane},'' {\em Nucl. Phys. B} {\bf 297} (1988)
  515--538.

\bibitem{Townsend:1995gp}
P.~K. Townsend, ``{P-Brane Democracy},''
  \href{http://www.arXiv.org/abs/hep-th/9507048}{{\tt hep-th/9507048}}.

\bibitem{Polchinski:1998rq}
J.~Polchinski, {\em String theory}.
\newblock Cambridge University Press, 1998.
\newblock Vol. 1: {A}n Introduction to the Bosonic String.

\bibitem{Green:1987sp}
M.~B. Green, J.~H. Schwarz, and E.~Witten, {\em Superstring Theory}.
\newblock Cambridge University Press, 1987.
\newblock Vol. 1: {I}ntroduction.

\bibitem{deWit:1988xki}
B.~de~Wit, M.~Luscher, and H.~Nicolai, ``{The Supermembrane Is Unstable},''
  {\em Nucl. Phys. B} {\bf 320} (1989) 135--159.

\bibitem{deWit:1989vb}
B.~de~Wit, U.~Marquard, and H.~Nicolai, ``{Area Preserving Diffeomorphisms and
  Supermembrane Lorentz Invariance},'' {\em Commun. Math. Phys.} {\bf 128}
  (1990) 39.

\bibitem{Bytsenko:1992pw}
A.~A. Bytsenko, K.~Kirsten, and S.~Zerbini, ``{Meinardus' theorem and the
  asymptotic form of quantum p-brane state density},'' {\em Phys. Lett. B} {\bf
  304} (1993) 235--238.

\bibitem{Bergshoeff:1995hm}
E.~Bergshoeff and E.~Sezgin, ``{Superp-Brane theories and new space-time
  superalgebras},'' {\em Phys. Lett. B} {\bf 354} (1995) 256--263,
  \href{http://www.arXiv.org/abs/hep-th/9504140}{{\tt hep-th/9504140}}.

\bibitem{Russo:1996if}
J.~G. Russo and A.~A. Tseytlin, ``{Waves, boosted branes and BPS states in m
  theory},'' {\em Nucl. Phys. B} {\bf 490} (1997) 121--144,
  \href{http://www.arXiv.org/abs/hep-th/9611047}{{\tt hep-th/9611047}}.

\bibitem{Russo:1996rw}
J.~G. Russo, ``{Stability of the quantum supermembrane in a manifold with
  boundary},'' {\em Phys. Lett. B} {\bf 392} (1997) 49--54,
  \href{http://www.arXiv.org/abs/hep-th/9609043}{{\tt hep-th/9609043}}.

\bibitem{Pavsic:1997eu}
M.~Pavsic, ``{The Dirac Nambu-Goto p-branes as particular solutions to a
  generalised, unconstrained theory},'' {\em Nuovo Cim. A} {\bf 110} (1997)
  369--396, \href{http://www.arXiv.org/abs/hep-th/9704154}{{\tt
  hep-th/9704154}}.

\bibitem{Yu:2011zza}
C.-X. Yu, C.~Huang, P.~Zhang, and Y.-C. Huang, ``{First and second quantisation
  theories of open p-brane and their spectra},'' {\em Phys. Lett. B} {\bf 697}
  (2011) 378--384.

\bibitem{Pavsic:2016icf}
M.~Pav\v{s}i\v{c}, ``{A New Approach to the Classical and Quantum Dynamics of
  Branes},'' {\em Int. J. Mod. Phys. A} {\bf 31} (2016), no.~20n21, 1650115,
  \href{http://www.arXiv.org/abs/1603.01405}{{\tt 1603.01405}}.

\bibitem{deWit:1988wri}
B.~de~Wit, J.~Hoppe, and H.~Nicolai, ``{On the Quantum Mechanics of
  Supermembranes},'' {\em Nucl. Phys. B} {\bf 305} (1988) 545.

\bibitem{Hoppe:1988gk}
J.~Hoppe, ``{Diffeomorphism Groups, quantisation and SU(infinity)},'' {\em Int.
  J. Mod. Phys. A} {\bf 4} (1989) 5235.

\bibitem{Taylor:2001vb}
W.~Taylor, ``{M(atrix) Theory: Matrix Quantum Mechanics as a Fundamental
  Theory},'' {\em Rev. Mod. Phys.} {\bf 73} (2001) 419--462,
  \href{http://www.arXiv.org/abs/hep-th/0101126}{{\tt hep-th/0101126}}.

\bibitem{Nambu:1973qe}
Y.~Nambu, ``{generalised Hamiltonian dynamics},'' {\em Phys. Rev. D} {\bf 7}
  (1973) 2405--2412.

\bibitem{Awata:1999dz}
H.~Awata, M.~Li, D.~Minic, and T.~Yoneya, ``{On the quantisation of Nambu
  brackets},'' {\em JHEP} {\bf 02} (2001) 013,
  \href{http://www.arXiv.org/abs/hep-th/9906248}{{\tt hep-th/9906248}}.

\bibitem{Hoppe:1996xp}
J.~Hoppe, ``{On M algebras, the quantisation of Nambu mechanics, and volume
  preserving diffeomorphisms},'' {\em Helv. Phys. Acta} {\bf 70} (1997)
  302--317, \href{http://www.arXiv.org/abs/hep-th/9602020}{{\tt
  hep-th/9602020}}.

\bibitem{Curtright:2002fd}
T.~Curtright and C.~K. Zachos, ``{Classical and quantum Nambu mechanics},''
  {\em Phys. Rev. D} {\bf 68} (2003) 085001,
  \href{http://www.arXiv.org/abs/hep-th/0212267}{{\tt hep-th/0212267}}.

\bibitem{Sheikh-Jabbari:2004fiz}
M.~M. Sheikh-Jabbari, ``{Tiny graviton matrix theory: DLCQ of IIB plane-wave
  string theory, a conjecture},'' {\em JHEP} {\bf 09} (2004) 017,
  \href{http://www.arXiv.org/abs/hep-th/0406214}{{\tt hep-th/0406214}}.

\bibitem{Isberg:1993av}
J.~Isberg, U.~Lindstrom, B.~Sundborg, and G.~Theodoridis, ``{Classical and
  quantised tensionless strings},'' {\em Nucl. Phys. B} {\bf 411} (1994)
  122--156, \href{http://www.arXiv.org/abs/hep-th/9307108}{{\tt
  hep-th/9307108}}.

\bibitem{Hassani:1994rf}
S.~Hassani, U.~Lindstrom, and R.~von Unge, ``{Classically equivalent actions
  for tensionless p-branes},'' {\em Class. Quant. Grav.} {\bf 11} (1994)
  L79--L85.

\bibitem{Bagchi:2015nca}
A.~Bagchi, S.~Chakrabortty, and P.~Parekh, ``{Tensionless Strings from
  Worldsheet Symmetries},'' {\em JHEP} {\bf 01} (2016) 158,
  \href{http://www.arXiv.org/abs/1507.04361}{{\tt 1507.04361}}.

\bibitem{Bagchi:2016yyf}
A.~Bagchi, S.~Chakrabortty, and P.~Parekh, ``{Tensionless Superstrings: View
  from the Worldsheet},'' {\em JHEP} {\bf 10} (2016) 113,
  \href{http://www.arXiv.org/abs/1606.09628}{{\tt 1606.09628}}.

\bibitem{Bagchi:2020fpr}
A.~Bagchi, A.~Banerjee, S.~Chakrabortty, S.~Dutta, and P.~Parekh, ``{A tale of
  three \textemdash{} tensionless strings and vacuum structure},'' {\em JHEP}
  {\bf 04} (2020) 061, \href{http://www.arXiv.org/abs/2001.00354}{{\tt
  2001.00354}}.

\bibitem{Bagchi:2021rfw}
A.~Bagchi, M.~Mandlik, and P.~Sharma, ``{Tensionless tales: vacua and critical
  dimensions},'' {\em JHEP} {\bf 08} (2021) 054,
  \href{http://www.arXiv.org/abs/2105.09682}{{\tt 2105.09682}}.

\bibitem{Bagchi:2019cay}
A.~Bagchi, A.~Banerjee, and P.~Parekh, ``{Tensionless Path from Closed to Open
  Strings},'' {\em Phys. Rev. Lett.} {\bf 123} (2019), no.~11, 111601,
  \href{http://www.arXiv.org/abs/1905.11732}{{\tt 1905.11732}}.

\bibitem{Bagchi:2020ats}
A.~Bagchi, A.~Banerjee, and S.~Chakrabortty, ``{Rindler Physics on the String
  Worldsheet},'' {\em Phys. Rev. Lett.} {\bf 126} (2021), no.~3, 031601,
  \href{http://www.arXiv.org/abs/2009.01408}{{\tt 2009.01408}}.

\bibitem{Bagchi:2022iqb}
A.~Bagchi, D.~Grumiller, and M.~M. Sheikh-Jabbari, ``{Horizon strings as 3D
  black hole microstates},'' {\em SciPost Phys.} {\bf 15} (2023), no.~5, 210,
  \href{http://www.arXiv.org/abs/2210.10794}{{\tt 2210.10794}}.

\bibitem{Bagchi:2023cfp}
A.~Bagchi, A.~Banerjee, J.~Hartong, E.~Have, K.~S. Kolekar, and M.~Mandlik,
  ``{Strings near black holes are Carrollian},'' {\em Phys. Rev. D} {\bf 110}
  (2024), no.~8, 086009, \href{http://www.arXiv.org/abs/2312.14240}{{\tt
  2312.14240}}.

\bibitem{Bagchi:2024rje}
A.~Bagchi, A.~Banerjee, J.~Hartong, E.~Have, and K.~S. Kolekar, ``{Strings near
  black holes are Carrollian. Part II},'' {\em JHEP} {\bf 11} (2024) 024,
  \href{http://www.arXiv.org/abs/2407.12911}{{\tt 2407.12911}}.

\bibitem{Bagchi:2024qsb}
A.~Bagchi, P.~Chakraborty, S.~Chakrabortty, S.~Fredenhagen, D.~Grumiller, and
  P.~Pandit, ``{Boundary Carrollian CFTs and Open Null Strings},''
  \href{http://www.arXiv.org/abs/2409.01094}{{\tt 2409.01094}}.

\bibitem{Gross:1987kza}
D.~J. Gross and P.~F. Mende, ``{The High-Energy Behavior of String Scattering
  Amplitudes},'' {\em Phys. Lett. B} {\bf 197} (1987) 129--134.

\bibitem{Gross:1987ar}
D.~J. Gross and P.~F. Mende, ``{String Theory Beyond the Planck Scale},'' {\em
  Nucl. Phys. B} {\bf 303} (1988) 407--454.

\bibitem{Pisarski:1982cn}
R.~D. Pisarski and O.~Alvarez, ``{Strings at Finite Temperature and
  Deconfinement},'' {\em Phys. Rev. D} {\bf 26} (1982) 3735.

\bibitem{Olesen:1985ej}
P.~Olesen, ``{Strings, Tachyons and Deconfinement},'' {\em Phys. Lett. B} {\bf
  160} (1985) 408--410.

\bibitem{Bagchi:2021ban}
A.~Bagchi, A.~Banerjee, S.~Chakrabortty, and R.~Chatterjee, ``{A Rindler road
  to Carrollian worldsheets},'' {\em JHEP} {\bf 04} (2022) 082,
  \href{http://www.arXiv.org/abs/2111.01172}{{\tt 2111.01172}}.

\bibitem{Bagchi:2024tyq}
A.~Bagchi, A.~Banerjee, I.~M. Rasulian, and M.~M. Sheikh-Jabbari, ``{Strings,
  Virasoro Sandwiches and Worldsheet Horizons},''
  \href{http://www.arXiv.org/abs/2409.16152}{{\tt 2409.16152}}.

\bibitem{Henneaux:1979vn}
M.~Henneaux, ``{Geometry of Zero Signature Space-times},'' {\em Bull. Soc.
  Math. Belg.} {\bf 31} (1979) 47--63.

\bibitem{Henneaux:2021yzg}
M.~Henneaux and P.~Salgado-Rebolledo, ``{Carroll contractions of
  Lorentz-invariant theories},'' {\em JHEP} {\bf 11} (2021) 180,
  \href{http://www.arXiv.org/abs/2109.06708}{{\tt 2109.06708}}.

\bibitem{Freidel:2022bai}
L.~Freidel and P.~Jai-akson, ``{Carrollian hydrodynamics from symmetries},''
  {\em Class. Quant. Grav.} {\bf 40} (2023), no.~5, 055009,
  \href{http://www.arXiv.org/abs/2209.03328}{{\tt 2209.03328}}.

\bibitem{Adami:2023wbe}
H.~Adami, A.~Parvizi, M.~M. Sheikh-Jabbari, V.~Taghiloo, and H.~Yavartanoo,
  ``{Carrollian structure of the null boundary solution space},'' {\em JHEP}
  {\bf 02} (2024) 073, \href{http://www.arXiv.org/abs/2311.03515}{{\tt
  2311.03515}}.

\bibitem{Grumiller:2019fmp}
D.~Grumiller, A.~PÃ©rez, M.~Sheikh-Jabbari, R.~Troncoso, and C.~Zwikel,
  ``{Spacetime structure near generic horizons and soft hair},'' {\em Phys.
  Rev. Lett.} {\bf 124} (2020), no.~4, 041601,
  \href{http://www.arXiv.org/abs/1908.09833}{{\tt 1908.09833}}.

\bibitem{Adami:2020amw}
H.~Adami, D.~Grumiller, S.~Sadeghian, M.~Sheikh-Jabbari, and C.~Zwikel,
  ``{T-Witts from the horizon},'' {\em JHEP} {\bf 04} (2020) 128,
  \href{http://www.arXiv.org/abs/2002.08346}{{\tt 2002.08346}}.

\bibitem{Adami:2021nnf}
H.~Adami, D.~Grumiller, M.~M. Sheikh-Jabbari, V.~Taghiloo, H.~Yavartanoo, and
  C.~Zwikel, ``{Null boundary phase space: slicings, news \& memory},'' {\em
  JHEP} {\bf 11} (2021) 155, \href{http://www.arXiv.org/abs/2110.04218}{{\tt
  2110.04218}}.

\bibitem{Freidel:2021fxf}
L.~Freidel, R.~Oliveri, D.~Pranzetti, and S.~Speziale, ``{The Weyl BMS group
  and Einstein\textquoteright{}s equations},'' {\em JHEP} {\bf 07} (2021) 170,
  \href{http://www.arXiv.org/abs/2104.05793}{{\tt 2104.05793}}.

\bibitem{Campiglia:2015yka}
M.~Campiglia and A.~Laddha, ``{New symmetries for the Gravitational
  S-matrix},'' {\em JHEP} {\bf 04} (2015) 076,
  \href{http://www.arXiv.org/abs/1502.02318}{{\tt 1502.02318}}.

\bibitem{Bagchi:2013bga}
A.~Bagchi, ``{Tensionless Strings and Galilean Conformal Algebra},'' {\em JHEP}
  {\bf 1305} (2013) 141,
\href{http://www.arXiv.org/abs/1303.0291}{{\tt 1303.0291}}.

\bibitem{Bars:1987dy}
I.~Bars, C.~N. Pope, and E.~Sezgin, ``{Massless Spectrum and Critical Dimension
  of the Supermembrane},'' {\em Phys. Lett. B} {\bf 198} (1987) 455--460.

\bibitem{Bars:1988uj}
I.~Bars, C.~N. Pope, and E.~Sezgin, ``{Central Extensions of Area Preserving
  Membrane Algebras},'' {\em Phys. Lett. B} {\bf 210} (1988) 85--91.

\bibitem{Dirac:1996}
P.~A.~M. Dirac, {\em {Lectures on Quantum Mechanics}}.
\newblock Belfer Graduate School of Science, Yeshiva University, New York,
  1996.

\bibitem{Thorne:1986iy}
K.~S. Thorne, R.~Price, and D.~Macdonald, {\em Black Holes: The Membrane
  Paradigm}.
\newblock Yale University Press,
1986.
\newblock

\bibitem{Price:1986yy}
R.~H. Price and K.~S. Thorne, ``{Membrane Viewpoint on Black Holes: Properties
  and Evolution of the Stretched Horizon},'' {\em Phys. Rev.} {\bf D33} (1986)
915--941.

\bibitem{Grumiller:2018scv}
D.~Grumiller and M.~M. Sheikh-Jabbari, ``{Membrane Paradigm from Near Horizon
  Soft Hair},'' {\em Int. J. Mod. Phys.} {\bf D27} (2018) 1847006,
\href{http://www.arXiv.org/abs/1805.11099}{{\tt 1805.11099}}.

\bibitem{Grumiller:2022qhx}
D.~Grumiller and M.~M. Sheikh-Jabbari, {\em {Black Hole Physics: From Collapse
  to Evaporation}}.
\newblock Grad.Texts Math. Springer, 11, 2022.

\bibitem{Dirac1950}
P.~A.~M. Dirac, ``generalised hamiltonian dynamics,'' {\em Canadian Journal of
  Mathematics} {\bf 2} (1950) 129-148.

\bibitem{Henneaux:1994lbw}
M.~Henneaux and C.~Teitelboim, {\em {quantisation of Gauge Systems}}.
\newblock Princeton University Press, 8, 1994.

\end{thebibliography}
\end{document}